# Multi-study integration of brain cancer transcriptomes reveals organ-level molecular signatures

Jaeyun Sung[1,2,*], Pan-Jun Kim[3,4], Shuyi Ma[1,2], Cory C. Funk[1], Andrew T. Magis[1,5], Yuliang Wang[1,2], Leroy Hood[1], Donald Geman[6,7], and Nathan D. Price[1]

**Authors' Affiliations:** [1]Institute for Systems Biology, Seattle, WA, USA, [2]Department of Chemical and Biomolecular Engineering, University of Illinois, Urbana, IL, USA, [3]Asia Pacific Center for Theoretical Physics, Pohang, Gyeongbuk, Republic of Korea, [4]Department of Physics, POSTECH, Pohang, Gyeongbuk, Republic of Korea, [5]Center for Biophysics and Computational Biology, University of Illinois, Urbana, IL, USA, [6]Institute for Computational Medicine, Johns Hopkins University, Baltimore, MD, USA, [7]Department of Applied Mathematics and Statistics, Johns Hopkins University, Baltimore, MD, USA.

*Current address: Asia Pacific Center for Theoretical Physics, Pohang, Gyeongbuk, Republic of Korea

**Corresponding author:**
Nathan D. Price, Institute for Systems Biology, E-mail: nprice@systemsbiology.org

## Abstract

We utilized abundant transcriptomic data for the primary classes of brain cancers to study the feasibility of separating all of these diseases simultaneously based on molecular data alone. These signatures were based on a new method reported herein – Identification of Structured Signatures and Classifiers (ISSAC) – that resulted in a brain cancer marker panel of 44 unique genes. Many of these genes have established relevance to the brain cancers examined herein, with others having known roles in cancer biology. Analyses on large-scale data from multiple sources must deal with significant challenges associated with heterogeneity between different published studies, for it was observed that the variation among individual studies often have a larger effect on the transcriptome than did phenotype differences, as is typical. For this reason, we restricted ourselves to studying only cases where we had at least two independent studies performed for each phenotype, and also reprocessed all the raw data from the studies using a unified pre-processing pipeline. We found that learning signatures across multiple datasets greatly enhanced reproducibility and accuracy in predictive performance on truly independent validation sets, even when keeping the size of the training set the same. This was most likely due to the meta-signature encompassing more of the heterogeneity across different sources and conditions, while amplifying signal from the repeated global characteristics of the phenotype. When molecular signatures of brain cancers were constructed from all currently available microarray data, 90% phenotype prediction accuracy, or the accuracy of identifying a particular brain cancer from the background of all phenotypes, was found. Looking forward, we discuss our approach in the context of the eventual development of organ-specific molecular signatures from peripheral fluids such as the blood.

## Introduction

One important goal in systems medicine is to develop molecular diagnostics that can accurately and comprehensively report health and disease states of an organ system [1,2]. The discovery of organ-level molecular signatures [3] from global biomolecule expression measurements would mark a significant advance toward this goal. In this regard, genome-wide transcriptomic data are readily available, making this a promising source for molecular signatures as well as a good means to study the robustness of signatures across different studies. During the past decade, transcriptomics analyses on clinical patient samples have been widely used to uncover cancer-associated genes [4] and to discover biomarkers for diagnosis, prognosis prediction, or optimal therapy selection [5-7]. Recently, RNAs measured in blood have also been used as serum-based molecular fingerprints of neurological disease [8].

While many molecular signature studies have focused on identifying differences between case (e.g., cancer) and control (e.g., normal), a more clinically relevant and challenging task is the multi-category classification problem. This task pertains especially to identifying signatures for molecular screening and monitoring purposes. Such signatures need to detect and stratify various pathological conditions simultaneously; they must therefore be highly specific for a particular disease as well as tissue of origin. The successful identification of more reliable and efficient molecular signatures will also be critical for the blood-based, organ-specific diagnostics envisioned for the future [9].

Data-driven, hierarchical approaches to multi-category classification have been investigated extensively in machine learning [10,11]. The basic idea of these methods is first to construct a classification framework in the form of a hierarchy, so that multi-category classifications can be reformulated into a series of binary decision sets (i.e., discriminating one class or group of classes from a second class or group of classes). The next step is to identify binary classifiers for all decision points (i.e., nodes and/or edges) of the hierarchy. This principle can be applied directly towards molecular disease classification, wherein all diseases can be organized into a global hierarchy of disease sets, where the diseases in each set share common expression patterns. The sets of binary classifiers can further be aggregated into a classifier marker-panel, which can direct diagnosis of an unlabeled patient sample down the hierarchical structure towards a particular label. Therefore, the cumulative expression patterns constitute "hierarchically-structured" molecular signatures.

A significant drawback to the use of molecular signatures derived from high-throughput—particularly transcriptomic—data is limited reproducibility and performance accuracy, which is often observed across independent studies of what are considered the same disease phenotype. This drawback holds true for both binary and multi-category classification problems. The lack of robustness, even for promising results, can be attributed to molecular heterogeneity within tumors or other diseased tissue-samples [12,13], complex disease subtypes, various patient demographics, and/or other biologically relevant factors. Another major issue is *batch effects*, which arise from differences or inconsistencies in experimental protocols, data quality, data-processing techniques, and laboratory conditions and personnel [14].

A promising method to address some of these limitations in robustness is to accumulate and combine data from many independent studies into large meta-analyses [15,16]. This integrated strategy naturally expands sample sizes across diverse sources and conditions and can therefore provide more reliable disease signatures as phenotype-associated signals become stronger relative to noise from batch effects and other sources of variance.

In this study, we developed a computational approach called Identification of Structured Signatures And Classifiers (ISSAC) to identify molecular signatures that simultaneously distinguish major cancers of the human brain. From an integrated dataset of publicly available gene expression data, ISSAC provides a global diagnostic hierarchy and corresponding structured brain cancer signatures composed of sets of gene-pair classifiers. The signal in the transcriptomics data was sufficient to develop accurate, comprehensive signatures, as long as the training set was sampled from the same population as the validation set (i.e., cross validation). In contrast, training on one dataset and testing against an independent set (i.e., an independent study measured from another lab) generally failed to reach the same performance due to biological and technical sources of dataset variation. To address this issue, we found that integration of datasets from multiple studies enhanced the disease signal sufficiently to mitigate batch effects and greatly improve independent validation results for brain cancers.

**Results/Discussion**

We compiled a multi-study, integrated dataset of brain cancer and normal transcriptomes [17-30] (Table 1, Table S1, and Table S2), on which we used our ISSAC algorithm (described below) to assemble classifiers into a node (Table 2, Table S3, and Figure 1) and a decision-tree (Table 3, Table S4, and Figure 2) marker panel. Importantly, while developing our algorithm to identify molecular signatures of brain cancer, we explored the effects of integrating data from multiple studies on classification performance, confirming that our integrated approach does indeed lead to more robust phenotype signatures.

Our marker panel consists of 39 total gene pairs and 44 unique genes (46 unique Affymetrix microarray probe IDs). Details on how the gene-pair sets were chosen as classifiers, and how they are used for phenotype prediction, can be found in the Materials and Methods section and Text S1. In addition, we discuss how the genes and gene pairs in our marker panel were found to have previously confirmed associations with brain cancer. Overall, we generated a marker panel with reasonably high multi-class brain cancer classification accuracy and straightforward biological interpretation.

**Table 1.** Description of all GEO microarray datasets used in this study.

| Phenotype name | GEO accession # | First author (publication year) | Ref. | Sample size | Affymetrix array |
|---|---|---|---|---|---|
| Ependymoma | GSE16155 | Donson (2009) | 17 | 19 | U133 plus2.0 |
| | GSE21687 | Johnson (2010) | 18 | 83 | U133 plus2.0 |
| Glioblastoma Multiforme | GSE 4412 | Freije (2004) | 19 | 59 | U133A |
| | GSE 4271 | Phillips (2006) | 20 | 76 | U133A |
| | GSE 8692 | Liu (2007) | 21 | 6 | U133A |
| | GSE 9171 | Wiedemeyer (2008) | 22 | 13 | U133 plus2.0 |
| | GSE 4290 | Sun (2006) | 23 | 77 | U133 plus2.0 |
| Medulloblastoma | GSE 10327 | Kool (2008) | 24 | 61 | U133 plus2.0 |
| | GSE 12992 | Fattet (2009) | 25 | 40 | U133 plus2.0 |
| Meningioma | GSE 4780 | Scheck (2006) | - | 62 | U133A/U133 plus2.0 |
| | GSE 9438 | Claus (2008) | 26 | 31 | U133 plus2.0 |
| | GSE 16581 | Lee (2010) | 27 | 68 | U133 plus2.0 |
| Oligodendroglioma | GSE 4412 | Freije (2004) | 19 | 11 | U133A |
| | GSE 4290 | Sun (2006) | 23 | 50 | U133 plus2.0 |
| Pilocytic Astrocytoma | GSE 12907 | Wong (2005) | 28 | 21 | U133A |
| | GSE 5675 | Sharma (2007) | 29 | 41 | U133 plus2.0 |
| Normal Brain | GSE 3526 | Roth (2006) | 30 | 146 | U133 plus2.0 |
| | GSE 7307 | Roth (2007) | - | 57 | U133 plus2.0 |

Studies that have not been published are denoted as '-'.

**Table 2.** The node marker-panel is a collection of gene-pair classifiers from the nodes of the diagnostic hierarchy.

| Node #[a] | Node classes[b] | Gene i[c] | Gene j[c] | k[d] |
|---|---|---|---|---|
| 2 | EPN GBM MDL MNG OLG PA | *PRPF40A* | *PURA* | 1 |
| 3 | normal | *PURA* | *PRPF40A* | 1 |
| 4 | EPN GBM MDL OLG PA | *NRCAM* | *ISLR* | 1 |
|  |  | *IDH2* | *GMDS* |  |
| 5 | MNG | *ISLR* | *NRCAM* | 1 |
| 6 | EPN GBM OLG PA | *SALL1* | *PAFAH1B3* | 2 |
|  |  | *SRI* | *NBEA* |  |
|  |  | *DDR1*[e] | *TIA1* |  |
|  |  | *DDR1*[e] | *MAB21L1* |  |
|  |  | *ITPKB* | *PDS5B* |  |
| 7 | MDL | *PAFAH1B3* | *SALL1* | 4 |
|  |  | *NBEA* | *SRI* |  |
|  |  | *TIA1* | *DDR1*[e] |  |
|  |  | *MAB21L1* | *DDR1*[e] |  |
|  |  | *PDS5B* | *ITPKB* |  |
| 8 | EPN | *NUP62CL* | *ZNF280A* | 2 |
|  |  | *GALNS* | *WAS* |  |
|  |  | *CELSR1* | *OR10H3* |  |
|  |  | *TLE4* | *OLIG2* |  |
| 9 | GBM OLG PA | *ZNF280A* | *NUP62CL* | 1 |
| 10 | GBM OLG | *DDX27* | *KCNMA1* | 1 |
|  |  | *COX7A2* | *GNPTAB* |  |
| 11 | PA | *KCNMA1* | *DDX27* | 3 |
|  |  | *GNPTAB* | *NDUFS2* |  |
|  |  | *APOD* | *PPIA* |  |
|  |  | *CD59* | *SNRPB2* |  |
|  |  | *SEMA3E* | *ADAMTS3* |  |
|  |  | *CD59* | *HINT1* |  |
|  |  | *BAMBI* | *CIAPIN1* |  |
| 12 | GBM | *FLNA* | *TNKS2* | 1 |
|  |  | *ITGB3BP* | *RB1CC1* |  |
|  |  | *DDX27* | *TRIM8* |  |
| 13 | OLG | *LARP5* | *ANXA1* | 1 |

[a]Node # corresponds to numerical labels in the diagnostic hierarchy shown in Figure 1. [b]Disease abbreviation (name): EPN (Ependymoma), GBM (Glioblastoma Multiforme), MDL (Medulloblastoma), MNG (Meningioma), OLG (Oligodendroglioma), PA (Pilocytic astrocytoma), and normal (Normal brain). [c]Gene i and gene j are the genes expressed higher and lower, respectively, within each gene-pair classification decision rule. Specifically, the statement of "Gene i is expressed higher than Gene j" being true contributes to the expression profile being classified as the phenotype(s) of the node. Gene names, chromosome loci, and Affymetrix microarray platform probe IDs of the classifier genes can be found in Table S1. [d]The minimum number of gene-pair classifiers whose decision rule outcomes for an expression profile are required to be 'true (= 1)' for the profile to be classified as the phenotype(s) of the node. [e]Genes that share same symbol/name, but correspond to different Affymetrix probe IDs.

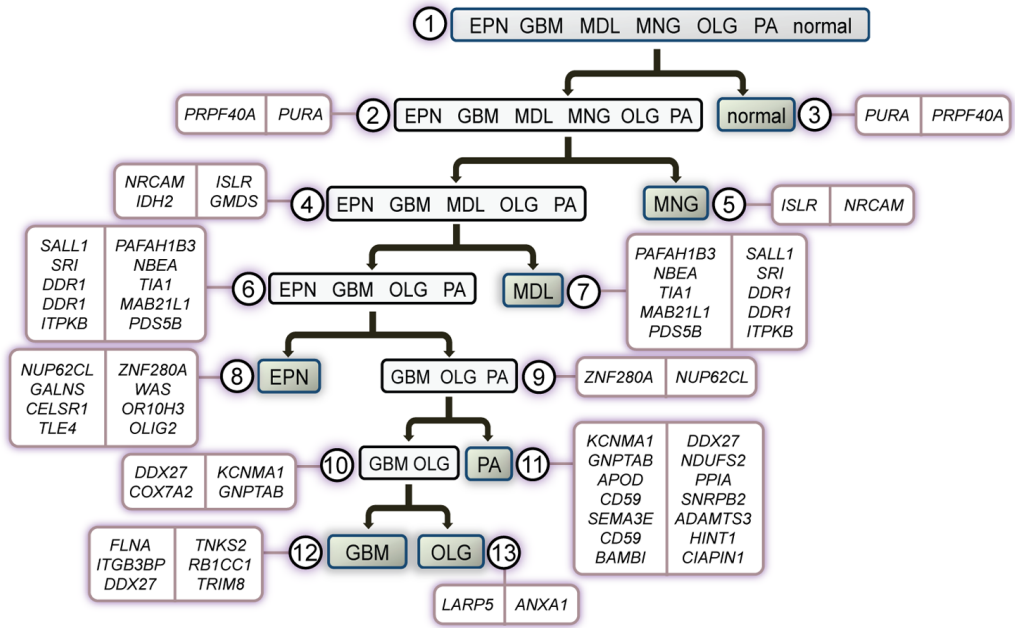

**Figure 1.** Gene-pair sets of the node marker-panel are shown at their corresponding twelve nodes in the brain cancer diagnostic hierarchy. Gene $i$ (left) and Gene $j$ (right) are the genes expressed higher and lower within each gene-pair, respectively. A transcriptome test sample is classified as the phenotype(s) of the node if the number of corresponding gene-pairs with a 'true' outcome for the statement "Gene $i$ is expressed higher than Gene $j$" is greater than or equal to a threshold $k$ defined for that node.

**Table 3.** The decision-tree marker-panel shows phenotype-specific signatures in the form of binary patterns.

| Gene symbols[a] | | Disease binary signatures[b] | | | | | | |
|---|---|---|---|---|---|---|---|---|
| Gene $i$ | Gene $j$ | EPN | GBM | MDL | MNG | OLG | PA | normal |
| *PRPF40A* | *PURA* | 1 | 1 | 1 | 1 | 1 | 1 | 0 |
| *NRCAM* | *ISLR* | 1 | 1 | 1 | 0 | 1 | 1 | - |
| *SRI* | *NBEA* | 1 | 1 | 0 | - | 1 | 1 | - |
| *NUP62CL* | *OR10H3* | 1 | 0 | - | - | 0 | 0 | - |
| *DDX27* | *KCNMA1* | - | 1 | - | - | 1 | 0 | - |
| *FLNA* | *TNKS2* | - | 1 | - | - | 0 | - | - |

[a]Affymetrix microarray platform probe IDs of the classifier genes are shown in Table S2. [b]For each gene-pair comparison (i.e., Is Gene $i$ > Gene $j$ ?), 1 and 0 delineates 'true' and 'false', respectively, and ' – ' denotes that the outcome is not used for classification.

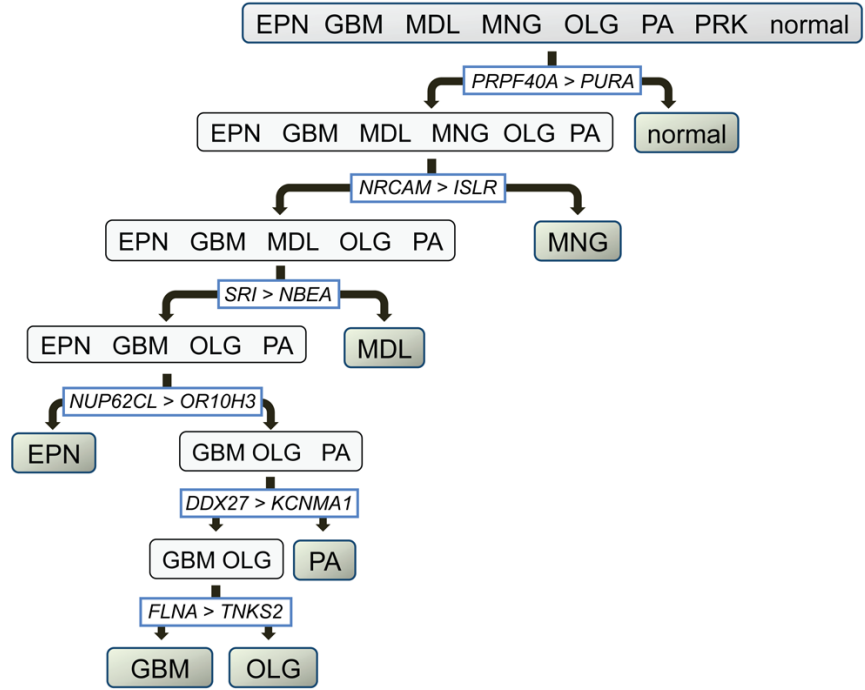

**Figure 2.** Gene-pairs of the decision-tree marker-panel are shown at their corresponding edges in the brain cancer diagnostic hierarchy. Gene *i* and Gene *j* are the genes expressed higher and lower within the gene-pair, respectively. For a given test sample, the direction of its classification down the diagnostic hierarchy is based on the gene-pair classifiers' true/false outcomes (left/right, respectively) for the statement "Gene *i* is expressed higher than Gene *j*".

**An overview of Identification of Structured Signatures And Classifiers (ISSAC)**

Here, we summarize the overall method of ISSAC into three main steps (Figure S1); a detailed algorithm and step-by-step guide are presented in the Materials and Methods section and Text S1, respectively. First, ISSAC constructs the framework for brain cancer diagnosis (Figure 3A and Figure S2)—a tree-structured hierarchy of all brain phenotypes including ependymoma (EPN), glioblastoma multiforme (GBM), medulloblastoma (MDL), meningioma (MNG), oligodendroglioma (OLG), pilocytic astrocytoma (PA), and normal brain, built using an agglomerative hierarchical clustering algorithm on gene expression training data. The construction of the hierarchy relies on iteratively identifying pairs of phenotype groups based on shared features in gene expression. As shown in Figure 3A, the cumulative set of different phenotypes is partitioned into smaller and more homogeneous subsets, thereby decomposing the multi-class diagnosis problem into more tractable sub-problems of class prediction.

Second, ISSAC identifies gene-pair classifiers corresponding to the nodes and edges of the diagnostic hierarchy (Figures 1 and 2 and Tables 2 and 3). Both types of classifiers are binary, i.e., attempt to distinguish between two sets of phenotypes. The objective of a node classifier is to distinguish the set of phenotypes associated with the node from *all other* phenotypes. For example, the classifiers of node 6 in Figure 1 and Table 2 can predict the class label of a particular transcriptome sample as either glioma (EPN, GBM, OLG, and PA) or non-glioma (MNG, MDL, and normal). In the case of an edge-based, decision-tree classifier, the objective is to distinguish the two sets of phenotypes associated with the two child nodes, analogous to rules of an ordinary decision tree. In the case of the two genes *PRPF40A* and *PURA* in Figure 2 and Table 3, this classifier determines the label of a sample as either brain cancer or normal phenotype. All classifiers are based on comparing the relative expression values (i.e., ranks) between two genes or several pairs of genes within a gene expression profile (Figures 1 and 2 and Tables 2 and 3). The chosen pairs are those that best differentiate between the phenotype sets and are based entirely on the *reversal of relative expression* (Materials and Methods), as previously reported [31]. Briefly, the decision rule by Geman *et al.* is based on two genes (e.g., gene *i* and gene *j*) for distinguishing between two phenotypes (e.g., class *A* and class *B*): If the expression of gene *i* is greater than that of gene *j* for a given profile, then the phenotype is classified as class *A*; otherwise, class *B*. It has been shown that using such simple decision rules with only a small number of gene pairs can lead to highly accurate supervised classification of human cancers [32,33]. We describe the advantages of using relative expression reversals in Text S2. In addition, we provide a summary of the expression differences between classifier genes *i* and *j* in Table S5.

Overall, the collection of node classifiers represent a series of coarse-grained to fine-grained explanations of the hierarchical groupings and are used in diagnosis to screen for phenotype-specific expression patterns (described below). Thus, the hierarchy of binary predictors guides classification of an expression profile in a dynamic *coarse-to-fine* fashion: a classifier is executed if and only if all of its ancestor classifiers have been executed and have returned a positive response—i.e., predicted the phenotypes in each node. The cumulative outcome of the node classifiers for a given expression profile is the set of its candidate phenotypes, corresponding to all the leaves of the hierarchy that were reached and tested positively. This property means that it is possible to traverse multiple paths to multiple leaf nodes, and thus multiple diagnoses may be made in this step (though in practice it is usually just one). For tie-breaking purposes, the decision-tree classifiers at the edges of the diagnostic hierarchy are used to reach a unique diagnosis.

Finally, ISSAC uses the gene-pair classifiers for class prediction (Figure 3B). Given a transcriptome sample, ISSAC executes the node classifiers in a hierarchical, top-down fashion within the disease diagnostic hierarchy to identify the phenotype(s) whose class-specific signature(s) is present. As shown in Figure 3B, transcriptome samples 4-7 all have expression signatures of at least one class, i.e., a sample is classified (positive) as at least one terminal node (leaf) phenotype. In contrast, samples 1-3 do not have any class-specific signatures, i.e., samples are not positive for any leaf, and are labeled as "Unclassified". In case of multiple class candidates, i.e., node classifiers for multiple leaves are positive as in samples 6 and 7, the ambiguity is resolved by aggregating all the decision-tree classifiers into a classification decision-tree, thereby leading any expression signature down one unique path toward a single phenotype. Once the hierarchy and classifiers were determined, ISSAC distinguished brain cancer phenotypes with an accuracy of 90% in ten-fold cross-validation (discussed below). When the individual transcriptomic samples used in the training set were re-examined, ISSAC correctly observed all samples with an apparent (resubstitution) accuracy of 94%. This gives a sense for the relatively small degree of over-fitting compared to the cross-validation accuracy estimate.

**A**

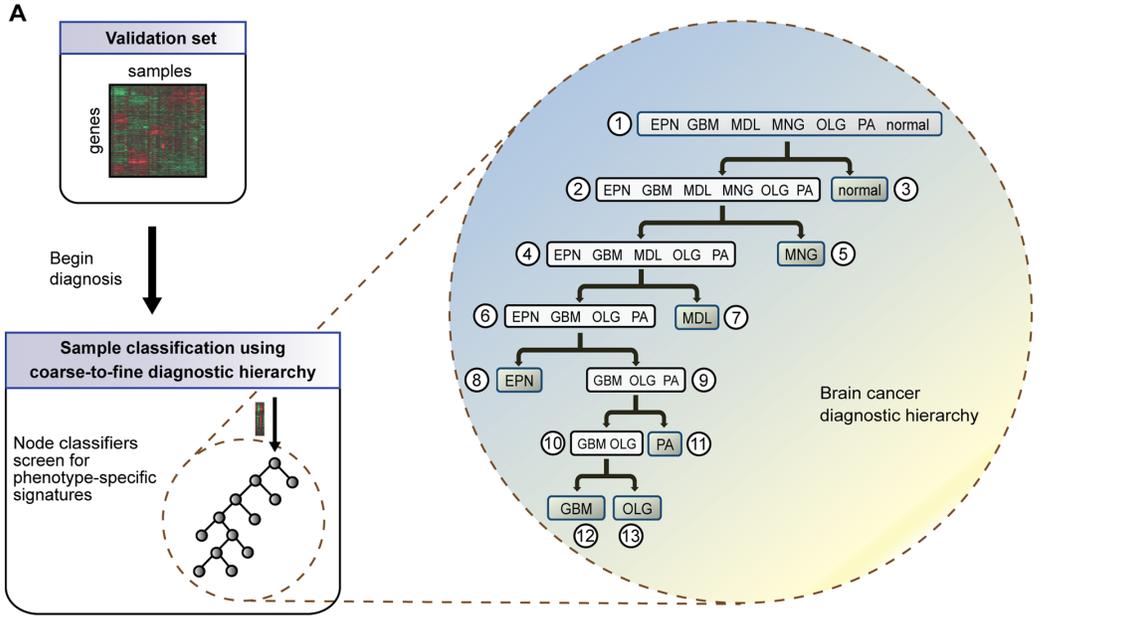

**B**

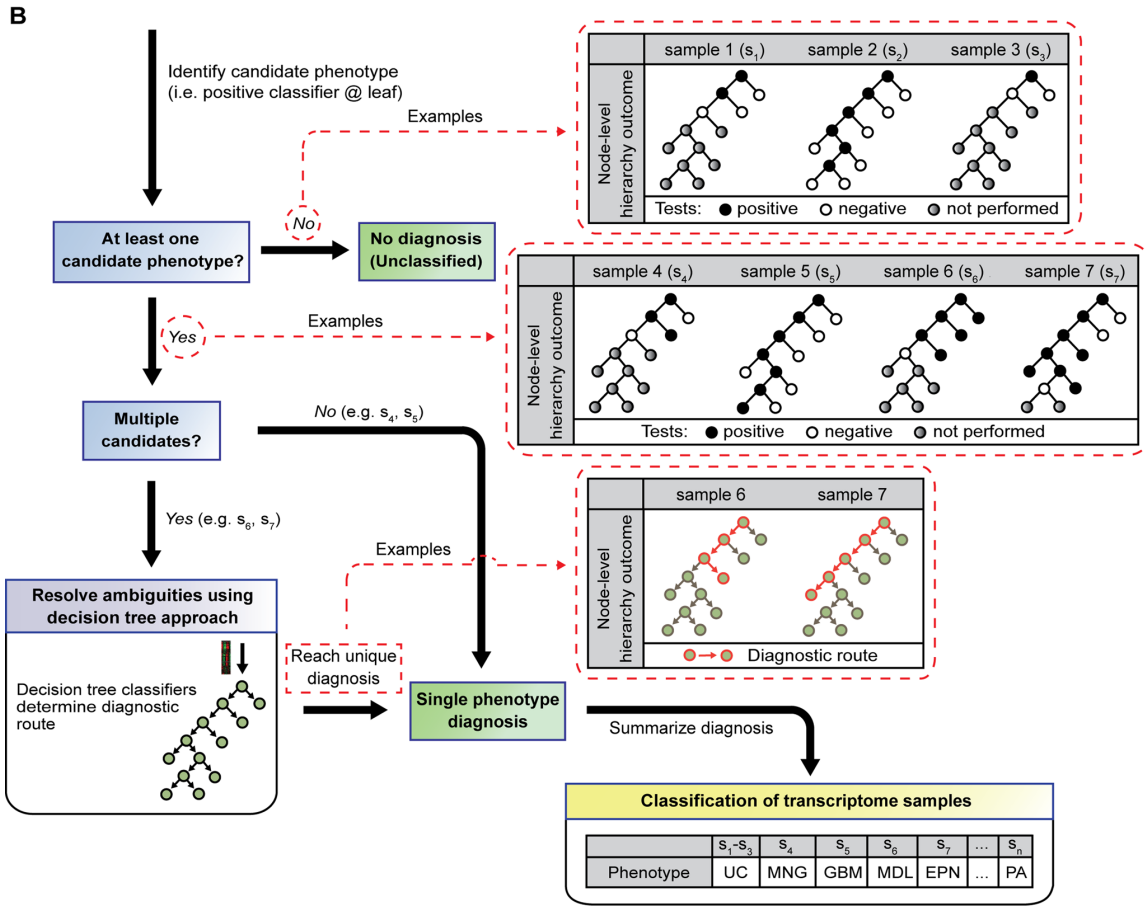

**Figure 3.** Comprehensive classification of human brain cancer and normal brain transcriptomes using molecular signatures from ISSAC. **A** The coarse-to-fine classification process is represented by a hierarchically structured groupings of phenotypes. There is a node classifier for each set of phenotypes in the hierarchy, which is designed to respond positively if the sample belongs to this set of diseases and negatively otherwise. Our diagnostic hierarchy has thirteen nodes in total, and seven terminal nodes (i.e., leaves). The node classifiers are executed sequentially and adaptively on a given expression profile; a classifier test for a particular node is performed if and only if all of its ancestor tests were performed and deemed positive. The node classifiers are used to screen for phenotype-specific signatures. **B** Leaves that have positive classifier outcomes correspond to the candidate phenotypes of a given expression profile. If there is no candidate phenotype, the expression profile is labeled as 'Unclassified'. If only one candidate phenotype is identified, the profile is labeled as that phenotype of the respective leaf. If the profile is considered to consist of multiple phenotype signatures, the ambiguity is resolved using the decision-tree classifiers based on the same diagnostic hierarchy. Here, the decision-tree classifiers are executed starting from the root of the tree, directing the profile to one of the two child nodes sequentially until it completes a full path towards a leaf. The phenotype label of the final destination corresponds to the unique diagnosis.

**Integrating disparate datasets identifies more robust molecular signatures across independent studies**

To estimate the robustness of signature accuracy, it is best to test molecular signatures against datasets (i.e., patient samples) that are truly independent of the training set (e.g., drawn from a different patient population, clinical laboratory, etc.). To study the effects of training across multiple studies, we used glioblastoma (GBM), where we had the highest number of transcriptomic datasets for the phenotype. We trained ISSAC on each of the five transcriptomic datasets (i.e., GSE #) of GBM, coupled in each case to all the data from the other brain phenotypes. The full multi-class signatures were completely relearned (every step) with the only difference in each case being which single GBM dataset was included in the training stage. We then assessed the accuracy of correctly classifying GBM transcriptomes measured in the four held-out datasets from all other possible phenotypes. We term this evaluation method as "hold-one-lab-in validation".

The overall hold-one-lab-in validation performance, or the average of all classification accuracies in Figure 3a, was 38%. This shows that, in general, individual datasets do not consistently yield robust molecular signatures. For example, GBM signatures from GSE8692 (6 samples, ref. 21) and GSE9171 (13 samples, ref. 22) led to average accuracies of 22% and 0% for classifying independent GBM samples from other studies, respectively. These significantly low performance results are not surprising for these sets given the very small sample numbers. To an extent, relatively larger datasets could indeed yield disease signatures of higher average accuracy. However, sample size was not the sole determining factor of signature performance. For example, training on GSE4412 (59 samples, ref. 19) gave an average accuracy of 23% (Figure 4a) on the remaining GBM samples from the other studies. As a notable exception, training on GSE4271 (76 samples, ref. 20) alone resulted in the best overall average accuracy (87%) in correctly classifying samples from the four held-out GBM datasets, with individual validation set accuracies ranging from 78% to 100% (Table S6). However, when GSE4290 (77 samples, ref. 23) was used as the training set, there was over a 30% lower average GBM classification accuracy (56%) despite the nearly identical sample size with GSE4271.

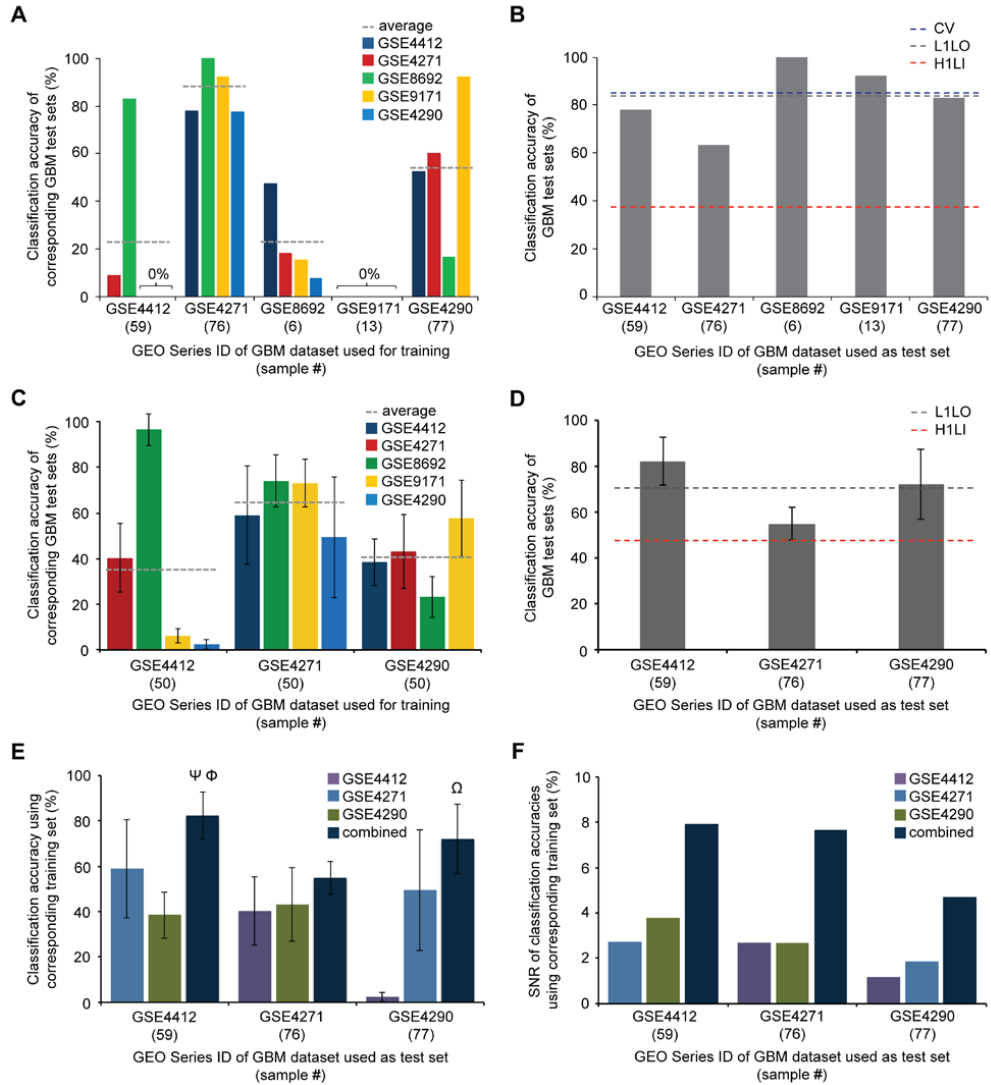

**Figure 4.** Molecular signatures from multi-study, integrated datasets have higher average phenotype prediction accuracy and lower performance variance than those from individual datasets. **A** Hold-one-lab-in validation results for each of the five glioblastoma (GBM) datasets. Gray line indicates average accuracy on the four validation sets. **B** Leave-one-lab-out validation results for each of the five GBM datasets. Blue and red line indicates average accuracy of GBM signatures from leave-one-lab-out (L1LO) validation and hold-one-lab-in (H1LI) validation, respectively. **C** H1L1 validation to test GBM signatures from GSE4412, GSE4271, and GSE4290, while keeping the number of samples in the GBM training set the same. 50 samples were randomly selected from each GBM dataset for signature learning. H1LI validation was executed ten times for each of the three GBM datasets. Error bars indicate standard deviations. **D** L1LO validation to test GBM signatures on GSE4412, GSE4271, and GSE4290 validation sets, while 50 total samples were randomly selected from the other four GBM datasets for signature learning. L1LO validation was executed ten times. **E** Data from **C** and **D** are used to show GBM signatures' accuracies on GSE4412, GSE4271, and GSE4290 validation sets when the GBM training data were from individual or combined GBM datasets. Ψ, Φ, and Ω indicate statistical significance relative to GSE4271, GSE4290, and GSE4412, respectively (Tukey's post-hoc test, $p < 0.01$). **F** Signal-to-noise ratios (SNRs) from data in **E**. SNR was calculated as the ratio of average accuracy to standard deviation.

We found considerable discrepancy between the minimum and maximum validation set accuracies for training sets GSE4412 (0% and 83%, respectively) and GSE4290 (17% and 92%) (Table S6). This indicates that batch effects, as well as potential biological discrepancies between populations studied at different sites, can lead to remarkable variation among transcriptomic datasets of supposedly the same phenotype. This "dataset variation" is widespread in large-scale expression studies, causing inconsistencies in molecular signature identification and performance reproducibility [34]. Large variation within and across transcriptomic datasets of GBM is perhaps not surprising, given that GBM is known to have various molecular subtypes [35]. Therefore, as mentioned above, molecular signatures from any single dataset need to be approached with caution in terms of their generalization.

We next analyzed how the multi-study integration approach affects performance robustness. One of each of the five datasets of GBM was sequentially withheld as the validation set, while all remaining gene expression data (including those from all other phenotypes) were used for training. The GBM signature was then evaluated on the held-out validation set. We term this strategy as "leave-one-lab-out validation". Classification accuracies using this approach ranged from 63% (GBM training set: 155 samples across four datasets; validation set: GSE4271, 76 samples) to 100% (GBM training set: 225 samples across four datasets; validation set: GSE8692, 6 samples) (Figure 4b). The average accuracy of the five leave-one-lab-out validations was 83%, which was considerably higher than that obtained from training on individual GBM datasets (38%). We conjecture that this result is due to the underlying variation in the training sets better representing the true variation in the population, both by achieving a greater sample size, as well as by having the samples come from a broader range of situations.

To evaluate how multi-study dataset integration alone affects performance robustness independent of sample size, we performed hold-one-lab-in and leave-one-lab-out validations for the studies with the largest number of samples, GSE4412, GSE4271, and GSE4290 (59, 76, and 77 samples, respectively) while training on the same number of samples for GBM. More specifically, the same steps in the analyses of Figure 4a and Figure 4b were used, while GBM signatures were learned from a training set of exactly 50 samples chosen randomly from either an individual dataset or across four combined datasets (with the fifth data set left out for validation). This process was conducted ten times for each GBM training set.

The average performances of hold-one-lab-in and leave-one-lab-out validations were 47% and 70%, respectively. Overall, the results were consistent with our two aforementioned conclusions: 1) when a molecular signature is learned from an individual dataset, its ability to accurately and precisely represent phenotype features across a broad population highly varies depending on the particular dataset used for training (Figure 4c and Table S7); and 2) combining datasets considerably increased average accuracy (Figure 4d and Table S7). Thus, dataset integration across multiple studies, even without change in sample size, can lead to significant improvements in predictive performance.

Lastly, we used the results in Figure 4c and Figure 4d to compare performances of different GBM signatures on the same validation set (Figure 4e). In all cases, signatures from combined datasets had, on average, higher classification accuracy than those from any of the individual datasets— even though the same number of samples was used in the training sets and were tested on a validation set independent of the training set. These results were then used to evaluate the precision of a GBM signature's classification accuracy by calculating its "signal-to-noise ratio (SNR)". SNR in the accuracy estimate was calculated herein as the ratio of average classification accuracy to standard deviation in the accuracy estimate across studies. We found that, for all validation set cases, GBM signatures developed on the basis of multiple datasets had SNRs greater by at least two fold than those from individual data sets. This clearly shows that learning on integrated datasets leads to molecular signatures that have higher and more consistent (i.e. less variable) predictive performance (Figure 4f), and motivated our choice in developing the brain cancer ISSAC signature to only use cases where we had at least 2 independent studies to learn across.

Overall, we have shown that when a broader range of conditions within a particular phenotype is presented during the classifier-learning stage, ISSAC can better distinguish the true disease signal from noise prior to independent validation. However, single and/or smaller training sets that were used to define the classifiers might not be representative of, or generalizable to, larger populations – leading to poor validation results. Therefore, the utilization of all currently available datasets from various sources and conditions may be a promising approach to finding novel diagnostic markers, and eventually bringing the successful adaptation of genomic biomarkers into clinical practice. Also, prospective design of studies is generally best when they utilize multiple sites to avoid over-fitting to particular contexts.

It is worth mentioning that in some cases, molecular signatures from a single source can have (or at least appear to have) superior performance, as demonstrated by the molecular signatures from GSE4271. Specifically, training on a single GSE4271 data set provided higher accuracy (87%, Figure 4a) than learning on any of the four sets combined (average 83%, Figure 4b). Indeed, when such surprisingly robust single datasets are identified, they potentiate significant new insight into the underlying heterogeneities present in a patient population of a disease phenotype. Such data sets can be utilized for follow-up studies, and hence serve as a valuable resource to the scientific and medical communities. It is, however, difficult in practice to predict in advance data set robustness, which must be ensured through careful sample collection and data set preprocessing techniques. To help ensure the production of reliable omics-based data sets, we recommend the following: 1) Good experimental design, such as clearly defining clinical phenotypes of interest; 2) When collecting new experimental data, sufficient sample size must be obtained; 3) All aspects of the experimental and analytical procedures must be carefully controlled to avoid batch effects; and 4) No confounding from factors unrelated to phenotype(s) of interest must occur.

**Brain cancer marker-panel achieves high classification accuracy in cross-validation**

As shown by our leave-one-lab-out validations, learning signatures across multiple datasets significantly improved average classification accuracy with concomitant reduction in performance variance. In this regard, the brain cancer marker-panel obtained using all currently available microarray data simultaneously (Tables 2 and 3) should represent more robust phenotype signatures.

The classification performance of this comprehensive brain cancer marker-panel was evaluated by ten-fold cross-validation (Figure S3). Our marker-panel achieved a 90% average of phenotype-specific classification accuracies (Table 4), showing strong promise against a multi-category, multi-dataset background at the gene expression level. In addition, we observed higher classification accuracy (93%) among the expression profiles for which a unique diagnosis was obtained without subsequent disambiguation from the decision-tree (Table S8). Furthermore, the glioblastoma (GBM) classification accuracy previously seen in our leave-one-lab-out analysis (83%) is comparable to that seen in cross-validation (85%). Indeed, that these two accuracies are so close suggests that, for GBM, the effects of variability among the datasets from different institutions and time-points have been mostly overcome by integration across multiple training studies.

**Table 4.** Classification performance of brain cancer marker-panel in ten-fold cross-validation.

| Actual phenotype | Predicted phenotype (%)[a] | | | | | | | | Total |
|---|---|---|---|---|---|---|---|---|---|
| | EPN | GBM | MDL | MNG | OLG | PA | normal | UC[b] | |
| EPN | **92.2** | 2.8 | 0.3 | 1.7 | 1.3 | 0.6 | 0.2 | 1.0 | 102 |
| GBM | 0.7 | **84.8** | 0.2 | 0.5 | 11.9 | 0.1 | 0.3 | 1.3 | 231 |
| MDL | 2.2 | 2.3 | **91.1** | 0.8 | 2.7 | 0.2 | 0.0 | 0.8 | 101 |
| MNG | 0.1 | 1.8 | 0.0 | **97.5** | 0.1 | 0.2 | 0.0 | 0.2 | 161 |
| OLG | 0.5 | 20.7 | 0.2 | 0.0 | **74.6** | 2.1 | 0.0 | 2.0 | 61 |
| PA | 1.3 | 2.3 | 0.0 | 0.0 | 1.3 | **94.4** | 0.0 | 0.8 | 62 |
| normal | 0.0 | 0.5 | 0.0 | 0.1 | 0.7 | 0.0 | **98.5** | 0.1 | 203 |

[a]Accuracies reflect average performance in ten-fold cross-validation conducted ten times. The main diagonal gives the average classification accuracy of each class (bold), and the off-diagonal elements show the erroneous predictions.
[b]UC (Unclassified samples). When using the node classifiers, expression profiles that did not exert a signature of any phenotype (i.e., did not percolate down to at least one positive terminal node) were rejected from classification. In this case, the Unclassified sample is treated as a misclassification.

Four other brain cancers (ependymoma, medulloblastoma, meningioma, and pilocytic astrocytoma) have estimated accuracies of at least 91%, suggesting clear differences between them and the other phenotypes at the transcriptomic level. The anatomical region specificity of these four cancers may have contributed toward their highly accurate separation, as there are regional areas of unique gene expression patterns. Roth *et al.* analyzed gene expression of 20 anatomically distinct regions of the central nervous system [30] and clustered all anatomical sites into distinct groups, providing evidence of region-specific expression patterns. However, results from another study analyzing gene expression data from distinct brain regions suggested that clustering disparities might also be due to activity of distinct brain cell types, rather than solely on region [36,37]. Furthermore, if region specificity played a dominant role in classification, we would expect to see a high number of misdiagnoses to occur between the normal brain, which was derived from 25 different locations (Text S3), and the six cancers. Such a trend was not observed in Table 4. Therefore, our predictive results suggest a stronger contribution from underlying cell-type specific and disease-intrinsic elements than from region effects alone.

Compared to the cross-validation accuracies of other phenotypes, lower performance was observed for GBM and oligodendroglioma (OLG) (85% and 75%, respectively). This could have been mainly a consequence of the limited ability of the marker-panel to correctly differentiate these two cancers from each other. Indeed, the distinction of these two phenotypes from transcriptomics seems to be rather difficult in general, and our accuracies here are comparable to those reported previously in two-phenotype comparison studies [38,39]. Furthermore, our signatures did show an excellent degree of sensitivity (96%) and specificity (97%) for distinguishing these two well-progressed gliomas as a set from all other brain phenotypes. There exist genetic tests and methods that differentiate GBM and OLG well, such as the combined loss of chromosome arms 1p and 19q [40], and over-expression of the transcription factor protein Olig2 [41], but our goal in this particular study was to evaluate molecular discriminatory power as represented in transcriptomes across multiple brain cancers.

**Marker-panel genes' association to cancer biology**

Several genes in our marker panel are strongly associated with brain cancers, suggesting putative relationships to the underlying pathophysiology of their corresponding phenotypes. One such gene is *NRCAM* (nodes 4 and 5 of Figure 1 and Table 2), which was reported as a marker for high-risk neuroblastoma [42] and poor prognostic ependymoma [43]. *NRCAM* was also found to be over-expressed in cell lines derived from pilocytic astrocytomas and glioblastoma multiforme tumors [44]. The receptor tyrosine kinase *DDR1*, a predicted marker gene for PA when expressed higher than *TIA1* and *MAB21L1* (nodes 6 and 7), was found to be over-expressed in high-grade gliomas and to promote tumor cell invasion [45]. *FLNA* was detected in the serum of high-grade astrocytoma (grade 3 and GBM) patients [46], and *ANXA1*, a gene that encodes an anti-inflammatory phospholipid binding protein, was implicated in astrocytoma progression [47]. These reports are consistent with our identification of both *FLNA* and *ANXA1* as two classifier genes expressed higher in GBM than in oligodendroglioma (nodes 12 and 13). The basic helix-loop-helix (bHLH) transcription factor *OLIG2* is innately expressed in oligodendrocytes and was recently characterized as a key antagonist of p53 function in neural stem cells and malignant gliomas [48]. In accordance with lower expression of *OLIG2* as an EPN classifier in this study (node 8), *OLIG2* expression was used as a negative marker to differentiate EPN from other gliomas [49]. *SEMA3E*, one of several classifier genes for PA (node 11), has been reported to drive invasiveness of melanoma cells in mice [50]. And finally, mutation to *IDH2* (node 4) in GBM is well known, with occurrence reported in 80% of secondary glioblastomas [51,52]. That the genes in our marker panel have previously confirmed ties to brain cancers raises the question of what is the underlying molecular framework surrounding the generation of gene-pair classifiers, which would be an interesting direction for future studies. Among the gene pairs in our marker panel, we focus on two pairs (below) in which the genes' common functional roles or relevance to cancer suggest putative relationships to corresponding pathology. Our discussions below point to potential biological relationships underlying the observed gene expression reversals, representing hypotheses that require further experimental validation.

One of the classifier gene pairs involved in the differentiation between meningioma and the remaining five brain cancers (EPN, GBM, MDL, OLG, PA) are two metabolic enzymes, *IDH2* and *GMDS* (node 4). *IDH2* converts isocitrate to α-ketoglutarate within the TCA cycle. This reaction produces NADPH, which not only is an essential cofactor for many metabolic reactions, but also helps to protect the cell against oxidative damage [53]. Moreover, *GMDS* aids the biosynthesis of GDP-fucose from GDP-mannose in mannose metabolism, in which NADPH is produced [54]. That the enzymatic activities of both *IDH2* and *GMDS* participate in the conversion between $NADP^+$ and NADPH is interesting, considering the well-known alteration to cellular metabolism and deregulated redox balance in cancer [55]. Possible MNG-specific mutations in *IDH2* and/or *GMDS*, or changes in the regulatory network that controls the expression of these two genes, may affect cellular redox balance and functions of other metabolic enzymes.

The *TLE4* and *OLIG2* gene pair is used to differentiate EPN from GBM, OLG, and PA (node 8). *TLE4*, a human homolog of the Drosophila Groucho protein, represses the Wnt and FGF developmental signaling pathways [56-58] by recruiting deacetylases to histones H3 and H4 [59]. FGF receptor signaling was reported to control neuronal and glial cell development by regulating *OLIG2* expression in zebrafish [58]. This connection between these two genes in regards to brain cell development could be reflective of the extent of cell-type differentiation (a hallmark of cancer), or lack thereof, unique to EPN compared with the other gliomas.

To develop further hypotheses of the functional relationships between the classifiers and pathophysiological traits, we looked for statistical enrichment of biological properties (e.g. biological processes, chromosome numbers) on an exhaustive list of gene-pairs discriminating GBM and OLG (Text S4). Our statistical enrichment of biological processes of gene-set *i* and gene-set *j* (the union of genes in each gene-pair classifier that are expressed relatively higher and lower in GBM, respectively) showed that the genes reflect disease properties. Specifically, the genes that are in gene-set *i*, or those expressed higher in GBM compared to OLG, were the most enriched in the biological process of 'Immunity and Defense' (Figure S4); this is in concordance with clinical observations showing high degree of inflammation inside malignant tumors (such as GBM), as well as the subsequent high number of immune cells. Our additional reports on the statistical enrichment of certain chromosome numbers link our classifiers to known genomic aberrations of their respective brain cancers, providing further insight as to why certain genes might have been selected as classifiers.

### Looking ahead: Molecular signatures based on putative blood borne biomolecules offer a glimpse into possible molecular diagnostics

The work reported herein has focused on identifying a structured molecular signature that can separate major brain cancers simultaneously, as well as on evaluating issues related to reproducibility in molecular signatures. However, our long-term motivation for wanting molecular signatures of an organ system is ultimately to find corresponding signatures in the blood, where they can be assayed non-invasively. Blood bathes virtually all organs, which secrete proteins and nucleic acids. Subsets of these secreted biomolecules can potentially constitute disease signatures for molecular diagnostics, as measurement technologies mature. Moreover, the blood is easily accessible in contrast to biopsies of diseased organs for obtaining transcript or protein profiles. In this regard, the brain represents an organ system where a critical need exists to develop non-invasive techniques to monitor its health state through secreted proteins.

Previously, organ-specific proteins have been detected in blood; when these proteins changed in concentration or chemical structure, the tissue origin of this change was identified [60]. For blood-based, organ-specific diagnostics, molecular signatures need to detect and stratify various possible cancers and other pathological conditions simultaneously. In the context of this current study, an intriguing question is if training ISSAC on shed or secreted blood borne biomolecule measurements identifies molecular signatures that allow us to distinguish health from disease; and if diseased, which one and how far has it progressed? Thus, the approach laid out herein for transcriptomics is a foundation for identifying similar signatures from blood proteins as these measurements become more abundant.

As proof of concept and to provide candidates for targeted proteomics analysis, we performed the above transcriptomic analysis of finding brain cancer signatures using only the genes that are annotated to encode extracellular proteins (Materials and Methods). We trained ISSAC on a total of 767 genes that matched this criterion, which led to a new brain cancer marker-panel composed of 41 gene-pair classifiers from 71 unique features (Figure S5). When looking at the case of GBM gene-pair classifiers, i.e. 59 node-based genes involved in the detection of either GBM or phenotype groups that include GBM, 11 were previously identified as potential GBM-specific serum markers (detected either from GBM cell-line secretome experiments or in human plasma): *APOD* [61], *CALU* [62], *CD163* [63,64], *CHI3L1* [65-67], *CSF1* [68,69], *EGFR* [68,70,71], *IGFBP2* [62,72-75], *NID1* [76], *PDGFC* [77,78], *PSG9* [72], and *PTN* [79]. We provide the functional roles of these genes in Table S9. None of these previous studies performed relative abundance comparisons or measured expression ratios, so we are unable to answer at this time whether the particular relative expression reversal patterns would be observed in serum. We were not able to find any direct available evidence associating the remaining GBM classifier genes to potential serum-based markers. Nevertheless, we were encouraged that ISSAC was able to verify some previously identified potential GBM markers, which provides support for its use towards a blood-based test since there is currently no clinically approved GBM-specific, serum-based biomarker.

Our marker-panel, composed entirely of genes encoding extracellular products, obtained an average classification accuracy of 87% in 10-fold cross-validation (Table S10), which compares favorably to the average accuracy we previously achieved using all the genes in the microarray (90%). This suggests that strong signal may possibly persist for phenotype distinction even when using only secreted biomolecules from diseased organs. If indeed there are enough biomolecules secreted into the blood at concentrations that can be accurately and consistently detected by e.g., targeted mass spectrometry, then there is the very exciting possibility that organ-specific pathologies, such as those described above, can be detected from the blood. This would truly make blood a powerful window into health and disease.

**Materials and Methods**

**Multi-study dataset of human brain cancer transcriptomes**

All transcriptomic data used in our analysis are publicly available at the NCBI Gene Expression Omnibus (GEO). We integrated 921 microarray samples of six brain cancers (ependymoma, glioblastoma multiforme, medulloblastoma, meningioma, oligodendroglioma, pilocytic astrocytoma) and normal brain across 16 independent studies into a transcriptome multi-study dataset. Importantly, we obtained the raw data (.CEL files) from each of these studies and preprocessed them uniformly using identical techniques to greatly reduce extraneous sources of technical artifacts (discussed below). All data manipulation and numerical calculations were performed using MATLAB (MathWorks).

To ensure data quality and to help control for systemic bias and batch effects), we used the following strict criteria and reasoning for brain phenotype selection: 1) Expression profiles must have been conducted on either the Affymetrix Human Genome U133A or U133 Plus 2.0 microarray platform. This allowed maximum microarray sample collection without considerable reduction in number of overlapping classifier features (i.e., microarray probe-sets). 2) Transcriptomic datasets (i.e., GSE #) for each phenotype must have been collected from at least two independent sources to help mitigate batch effects. 3) All datasets must have consisted of no fewer than 5 microarray samples. 4) All datasets must have originated from primary brain tumor or tissue biopsies. Expression profiles from cell-lines or laser micro-dissections were not used in our study to better ensure sample consistency. 5) Raw microarray intensity data (.CEL files) must have been available on GEO for consensus preprocessing (described below). 6) Sample preparation protocols must have been fully disclosed on GEO. 7) All microarray samples in a dataset of a given phenotype were used in order to take into consideration all sources of heterogeneity. That is, *no* samples were excluded because their gene expression profiles were abnormal for their associated phenotypes. We are aware that this may allow mislabeled samples, e.g. samples that were originally misclassified by the histopathologist upon class labeling (Text S5), to be used in the classifier-learning stage, and thereby limit the biological "purity" of a phenotype in the training set. This can pose a serious challenge in interpreting misclassified samples that actually seem to be a much better match (or even perfect match) to a different phenotype, leading to questions of whether a misclassification is due to ISSAC's limitation in distinguishing phenotypes, or whether a re-evaluation of the original tumor biopsy is required. Despite these concerns, we concluded this to be the most stringent test. After an exhaustive search on GEO, we identified 921 microarray samples from 16 studies that met the above criteria (as of January 2011). Information on all datasets (e.g., publication sources, Affymetrix platforms, GEO dataset IDs, and microarray sample IDs), studies, and GEO microarray sample IDs used in our study is available in Table 1, Table S1, and Table S2, respectively.

Raw microarray intensity data (.CEL files) were obtained online from GEO and preprocessed uniformly. More specifically, common probe-sets were found across all transcriptome samples, and consensus preprocessing was performed on all the raw microarray image data to build a consensus dataset. This step removes one major non-biological source of variance between different studies. These preprocessed samples were used to build a multi-study integrated dataset of human brain cancer and normal brain transcriptomes. Finally, stringent probe-set filtering was used to remove spurious classifier features. Our consensus preprocessing and probe-set filtering methods are explained in further detail below. Our integrated and uniformly pre-processed dataset is available on our group's webpage (http://price.systemsbiology.net/downloads) as a community resource for those who wish to conduct their own analyses.

**Consensus preprocessing using GCRMA**

All gene expression data used in our study were measurements conducted on either the Affymetrix Human Genome U133A or U133Plus2.0 oligonucleotide microarrays. The expression level of a target gene on these two platforms is measured by first quantifying the total intensity of fluorescently labeled RNA fragments (from patient specimens) that bind to a probe set, or the set of complementary 25-mer oligonucleotide probe sequences. The intensities of all probe sets (raw measurements in the form of .CEL files) are then adjusted for background variability and normalized across all samples to obtain the target genes' final expression values.

Raw .CEL data files were downloaded directly from GEO. Probe set information used in this study were based on the latest Affymetrix annotations. Raw intensity measurements of all microarray samples considered in our study were preprocessed collectively (consensus preprocessing) using the MATLAB implementation of the microarray preprocessing GCRMA [80]. Only the probe sets that map to known genes and exist on both Affymetrix platforms (same oligonucleotide sequences) were considered for preprocessing. The use of individual Affymetrix probe sets as classifiers (and not the mean or median of their expression values as demonstrated in other microarray-based studies) imposes limitations in the classifiers' multi-platform compliance, as discussed in Text S6 and Text S7.

**Probe set filtering using MAS5 calls**

Probe sets of Affymetrix microarrays have "perfect match" probes that are exactly complementary to the target gene's mRNA sequence. They also have "mismatch" probes that contain a mismatched nucleotide halfway along the probe sequence, and are used to estimate the degree of non-specific binding. To ensure that a probe set is reliably detected, the measurements of the "perfect match" probes must be significantly greater than those of the "mismatch" probes. This is usually assessed based on statistical measures. The MAS5 preprocessing software makes expression quality calls based on the nonparametric Wilcoxon signed-rank test. The "absent" call is made when the $p$-value is greater than 0.06, representing no significant difference between the measurements of the "perfect match" and those of the 'mismatch' probes [81]. We eliminated probes that were determined to be "absent" in all samples of the consensus dataset. After this probe set filtering step, 19,656 probe sets (corresponding to target genes) within each microarray sample were kept for further analysis.

All GCRMA preprocessing and MAS5 probe set filtering procedures were conducted separately for training and test set samples, i.e., inside each cross-validation or hold-out loop, in order to avoid possible cross-talk between the two datasets. Genes that were excluded based on the MAS5 "absent" calls on the training data were also removed from the corresponding test data.

**Description of ISSAC algorithm**

**A tree-structured framework for brain cancer diagnostics**

Let $\mathcal{L}$ denote the set of class labels, in our case the seven brain phenotypes: six cancers and normal. Given an expression profile $\mathbf{x}$, the objective is to determine its true phenotype $Y \in \mathcal{L}$. The main assumption is that there are natural groupings $L \subset \mathcal{L}$ among the phenotypes. Thus, testing for these groupings can more efficiently utilize the available training data, leading to more accurate classification than testing for each phenotype individually. Based on these attributes, the natural structure to represent $\mathcal{L}$ is then a diagnostic hierarchy in the form of a binary hierarchical decision tree $T$. Each node $t \in T$ is associated with a set of phenotypes $L_t \subset \mathcal{L}$. The root of $T$ contains all the phenotypes and each leaf (terminal node) of $T$ delineates a single phenotype. Overall, this representation is nested, in the sense that the set of phenotypes at every non-terminal node is the disjoint union of the phenotypes of the two child nodes. This tree is built from the training data by agglomerative hierarchical clustering derived from features of the profiles, as discussed below.

**Node classifiers are assembled according to the diagnostic hierarchy**
There is a binary classifier $f_t$ for every node $t \in T$ except for the root. The classifier $f_t$ is a function of the expression profile $\mathbf{x}$. Put simply, $f_t$ is a collective "test" for phenotypes in $L_t$ versus all other phenotypes. More formally, the classifier returns two possible outcomes: $f_t(\mathbf{x}) = 1$ (i.e., positive) signals that we accept the hypothesis that $Y \in L_t$, and $f_t(\mathbf{x}) = 0$ (i.e., negative) signals that we reject this hypothesis and conclude that $Y \notin L_t$. In particular, $f_t$ is *not* a test for $L_t$ versus the phenotypes in the sibling of $t$, as would be the case with a standard decision tree. Rather, $f_t$ looks for traits within a given profile $\mathbf{x}$ which characterize all phenotypes in $L_t$ *simultaneously*, such that a positive result signifies that the classifier assumes the true class of $\mathbf{x}$ belongs to $L_t$.

Classifier learning begins at the two child nodes of the root, and the classifiers are learned from two types of training data. The positive training data for learning the classifier $f_t$ for node $t$ are all the expression profiles of the phenotypes in $L_t$. The negative training data are all the profiles of the phenotypes that are not in $L_t$.

Being binary, each classifier has two performance metrics: the *sensitivity* of $f_t$ is the probability that $f_t(\mathbf{x}) = 1$ given $\mathbf{x}$ is from the positive training data, and the *specificity* of $f_t$ is the probability that $f_t(\mathbf{x}) = 0$ given $\mathbf{x}$ is from the negative training data. Due to the coarse-to-fine, hierarchical manner in which the classifiers are processed, we required the *sensitivity* of $f_t$ to be as close to unity as possible. This can be accomplished at the expense of specificity by adjusting a threshold, as discussed below. The reason for imposing a high sensitivity on each classifier is that if a test profile is rejected from belonging to $L_t$ by the classifier when in fact it does belong to $L_t$, it cannot be recovered. However, the reduced specificity is only local to each node, and the overall specificity increases with testing at subsequent nodes.

**A coarse-to-fine screening yields candidate phenotypes**
The strategy for processing any given profile $\mathbf{x}$ with the diagnostic hierarchy is breadth-first, coarse-to-fine. Starting from the two child nodes of the root, classifiers are executed sequentially and adaptively, with $f_t$ performed if and only if all its ancestor tests are performed and are positive. More specifically, $f_t$ is performed if and only if $f_s = 1$ for every node $s \in T$ between $t$ and the root. As soon as $f_t = 0$ for a non-terminal node $t$, none of the descendant classifiers in the sub-tree rooted at $t$ are performed. This is because a negative response of $f_t$ means that the phenotype is unlikely to belong to $L_t$ and the set of phenotypes associated with descendant of $t$, which are necessarily subsets of $L_t$. This facilitates pruning whole subsets of phenotypes at once.

The complete coarse-to-fine screening process for $\mathbf{x}$ results in a *set* of detected phenotypes. We denote this set by $L(\mathbf{x}) \subset \mathcal{L}$. These are the phenotypes corresponding to a complete chain of positive results for all $f_t$ from root to leaf. Equivalently, $L(\mathbf{x})$ is the set of phenotypes that are not ruled out by any test performed. During the diagnostic process, a profile may traverse only one path all the way to the terminal node. In this case, $L(\mathbf{x})$ consists of a single phenotype $d$, and the diagnostic process terminates with $Y = d$ as the predicted phenotype. However, a profile may also traverse multiple branches to the terminal nodes of $T$, in which case $L(\mathbf{x})$ consists of multiple candidate phenotypes (see the discussion on resolving ambiguities below). Moreover, a profile may reach no terminal nodes, in which case $L(\mathbf{x})$ is empty. When no terminal node is reached, the profile is determined to be outside of $L$, and labeled as 'Unclassified'.

**Resolving ambiguities using a decision-tree approach**

When $L(\mathbf{x})$ consists of multiple phenotypes, it becomes necessary to refine the diagnosis. The ambiguity is resolved by another tree-structured process – an ordinary decision tree based on the same diagnostic hierarchy. For every pair of sibling nodes, $L_t = \{EPN\}$ and $L_s = \{GBM, OLG, PA\}$, we learn a classifier $g_{t,s}$ which tests $Y \in L_t$ versus $Y \in L_s$, just as in an ordinary decision-tree (the process of classifier identification is elaborated below). Starting from the root of the tree, execution of the decision-tree classifiers directs a profile to one of two sibling nodes sequentially until it reaches a terminal node. Unlike the process of traversing the hierarchy of node classifiers, a sample that enters the decision tree is directed to one and only one leaf node, and hence uniquely labeled.

**Classifier design and learning**

Every node classifier is based on expression level comparisons between two genes. Let $G$ be the set of all genes for which we have microarray expression data, and denote the set of all distinct *pairs* of genes by $\mathcal{P}$. For each gene-pair $\left(g_i, g_j\right) \in \mathcal{P}$, consider the Boolean feature $Z_{ij}(\mathbf{x}) \in \{0,1\}$ of an expression profile $\mathbf{x} = \left\{x_g, g \in G\right\}$. $Z_{ij}(\mathbf{x})$ assumes the value 1 if gene $g_i$ is expressed higher than gene $g_j$ (i.e., $x_{g_i} > x_{g_j}$) in $\mathbf{x}$, and the value 0 otherwise (i.e., $x_{g_i} \leq x_{g_j}$). These are the features that have been used in previous work on relative expression reversals [31]. Each node classifier $f$ is constructed from a small set of gene-pairs $P \subset \mathcal{P}$, the binary outcomes of $Z_{ij}$ for all $(i,j) \in P$, and a constant threshold $k$. More specifically, $f(\mathbf{x}) = 1$ if $\sum_{\left(g_i, g_j \in P\right)} Z_{ij} \geq k$, and $f(\mathbf{x}) = 0$ otherwise. The threshold $k$ takes values between 1 and $|P|$.

There is a classifier of this nature for every node $t \in T$, except for the root. The set of gene-pairs $P = P_t$ and threshold $k = k_t$ depend on the node $t$. Hence, for each $t$, the classifier $f_t = 1$ if at least $k_t$ of the gene-pair comparisons in $P_t$ are positive $\left(Z_{ij} = 1\right)$; otherwise, $f_t = 0$. The comparisons are chosen such that, for each pair $\left(g_i, g_j\right)$ in $P_t$, we expect to see gene $g_i$ expressed more than gene $g_j$ under the assumption that the phenotype of $\mathbf{x}$ belongs to $L_t$, whereas if the phenotype of $\mathbf{x}$ does not belong to $L_t$, we expect to see the reverse. For every node $t$, every pair of all gene-pair combinations is "scored" by the difference between the probability of the event that $Z_{ij} = 1$ given $Y \in L_t$ and the probability given $Y \notin L_t$. These probabilities are estimated from the training data, and the subset of pairs with the highest scores are chosen.

Since each positive (resp., negative) comparison is viewed as evidence for $Y \in L_t$ (resp., $Y \notin L_t$), we can then favor sensitivity over specificity by varying the threshold $k_t$. That is, by choosing a relatively small value for $k_t$ relative to the number of comparisons in $P_t$, we can make it highly likely that the classifier responds positively when in fact the sample belongs to the set $L_t$. We show the sets of gene-pairs $P_t$ for each of the twelve nodes in our diagnostic hierarchy in Table 2 and an illustrative example in Figure 1. Finally, the decision tree classifiers $g_{t,s}$ are all based on comparisons of *single* gene-pairs at all edges of the diagnostic hierarchy.

While multiple gene pairs were used at each decision point in the node-based tree, only a single gene pair was used at each decision point in the decision-tree. This is due to the difference in the motivation of building the two trees; the node-based tree was constructed to maximize sensitivity and minimize false-positives with as many pairs as necessary, while the decision-tree was designed to resolve multiple diagnoses (i.e. ties) which could be done with only one pair. We show the pair of genes for each of the six decision-tree classifiers in Table 3 and Figure 2. MATLAB implementations of the ISSAC algorithm and a step-by-step tutorial are available to download at http://price.systemsbiology.net/downloads.

**Selecting genes that encode extracellular products**
Using Gene Ontology (GO) annotations, we have identified a list of 767 genes (mapped on 1,085 total probes) in every transcriptome sample that encode for possible blood-borne proteins. Specifically, we selected only the genes whose products are annotated to be in either the 'Extracellular Space' or the 'Extracellular Region' cellular locations. We use this gene set as a starting point for targeted blood diagnostics. All computational steps and analyses in regards to molecular signature discovery are identical to those discussed above.

**Acknowledgements**

We would like to thank James A. Eddy and Matthew C. Gonnerman for helpful discussions.

# Supplementary Text

## Text S1. Step-by-step description of how ISSAC works

### a. Construction of the disease diagnostic hierarchy

Let L = $(d_1, \ldots, d_7)$ be the collection of class labels, where $d_i$ denotes brain phenotype $i$. Using expression profiles of the phenotype classes, we first calculate the Top Scoring Pair (TSP) score ($\Delta$) of all gene-pair combinations between all pair-wise class comparisons. As previously described [1], the TSP score between two classes $d_m$ and $d_n$, of two genes, gene $i$ and gene $j$, is defined as:

$$\Delta_{i,j}(d_m, d_n) \ = \ \left| p_{i>j}(d_m) - p_{i>j}(d_n) \right| ,$$

where $p_{i>j}(d_m)$ and $p_{i>j}(d_n)$ denotes the percentage of samples in $d_m$ and $d_n$, respectively, whose expression of gene $i$ is higher than that of gene $j$. $\Delta_{max}(d_m, d_n)$ denotes the maximum $\Delta_{i,j}$ between $d_m$ and $d_n$ over all gene pairs $i$ and $j$

Let $C$ designate an evolving set of groups of labels that starts off as the set of individual class $(d_1, \ldots, d_7)$. The brain disease diagnostic hierarchy was constructed by progressively evolving $C$ towards the set of all groupings in the hierarchy using the following steps:

1. For all pair-wise comparisons of distinct elements in $C$, we calculate all $\Delta_{max}$. The leaves of the class-pair $d_m$ and $d_n$ with the smallest value of $\Delta_{max}$ are merged into the first node of the tree, denoted as $n_{d_m, d_n}$.

2. $\Delta_{max}$ of all pair-wise comparisons of the elements in the updated $C$ are calculated, and the pair with the smallest value of $\Delta_{max}$ is grouped into the next node of the tree. Since at this point $C$ contains one non-singleton node and a host of other leaves, the next merging can be either between two leaves $d_u$ and $d_v$, denoted as $n_{d_u, d_v}$, or between a node $n_{d_m, d_n}$ and a leaf $d_u$, denoted as $n_{d_m, d_n, d_u}$. Whichever pair with the smallest $\Delta_{max}$ merges to form a new node in $C$.

3. This process of finding the minimum $\Delta_{max}$ for all pair-wise elements in $C$, and adding the new node in $C$, is iterated until all nodes and leaves are connected to form a tree structure. All classes combine to form the top node $n_{d_1, \ldots, d_7}$ at the top of the diagnostic hierarchy (i.e. root).

### b. Identification of the node marker panel

The general idea of the node marker panel discovery method is to find a classifier at every node (excluding the root) and leaf of the *diagnostic hierarchy*. The node classifiers are based on common expression attributes of phenotypes grouped within a particular node; *these classifiers consist of a set of gene-pair binary decision rules*, whose collective 'true (= 1)' or 'false (= 0)' outcomes are to guide classification of a transcriptome test sample towards a brain phenotype. The gene-pair classifiers of the node marker panel are identified through the following steps:

1. Let $X_t$ denote all samples included in the training of node $t$. These samples correspond to phenotypes of either $Y_t \in L_t$ or $Y_t \in L_{t^c}$, where $L_t$ is the subgrouping of classes at $t$ (we start from a child nodes of the root), and $L_{t^c}$ represents the set of classes that do not belong to $L_t$. Using all expression profiles in $X_t$, we identify nine disjoint gene pairs $(g_i, g_j)$ with the nine highest values of $\Delta_{i,j}$ between $L_t$ and $L_{t^c}$. This set of gene pairs is denoted as $P_t = (p_1, \ldots, p_9)$, where $p_m$ is the gene-pair with the $m^{th}$ highest $\Delta_{i,j}$. Here, for any given transcriptome sample, the two genes of $p_m$ comprise the following decision rule: If a sample displays the relative expression relation $x_{g_i} > x_{g_j}$, then the sample is classified as $L_t$; otherwise, the sample is classified as $L_{t^c}$, where $x_{g_i}$ and $x_{g_j}$ are expression levels of gene $i$ and gene $j$, respectively.

When multiple gene pairs achieve the same $\Delta_{i,j}$, the gene pairs are preferentially selected by the tie-breaking scheme employed by Tan *et al.* [1]. Note that since the classification compares $L_t$ and $L_{t^c}$ at each node $t$, the classifiers of the two child nodes of the root must necessarily be the same. This special case occurs in the child nodes of the root because every class label is represented either in the left child or right child.

2. For the $n$ gene pairs in $P_t$ with the $n$ highest TSP scores ($x = 1, \ldots, 9$), a constant threshold $k$ ($k \leq n$) is found, representing the minimum number of gene pairs required to have 'true ($= 1$)' decision rule outcomes in order to have a particular sample classified as $L_t$. The optimal $n$ gene pairs and threshold $k$ are found concurrently: $n$ is the fewest number of gene pairs that yields classification sensitivity (percentage of $X_t$ samples that are classified as $Y_t \in L_t$ correctly) above a desired value; $k$ is optimized to be the threshold that results in the highest overall accuracy, i.e. percentage of all samples in $X_t$ that are classified correctly.

3. Using the optimal $n$ gene pairs and threshold $k$ found in Step 2, we evaluate all $n$ gene-pair classifiers on every sample of $X_t$, where each classifier's binary decision rule outcome (on each sample) is either 'true ($= 1$)' or 'false ($= 0$)'. The total number of true outcomes for each sample is denoted as $\mu$. If $\mu \geq k$, the sample is labeled 'positive' for $L_t$ and passes into $X_{t^*}$ of child node $t^*$ of $t$ for further classifier training. However, if $\mu < k$, the sample is labeled 'negative' for $L_t$ and does not pass into $X_{t^*}$. In general, most samples of $Y_t \in L_t$ passes into $X_{t^*}$, while most samples of $Y_t \in L_{t^c}$ do not pass into $X_{t^*}$. This updated collection of samples of $X_{t^*}$ is now to be used in Step 4 by the child nodes $t^*$.

4. We iterate Steps 1-3 on the sibling node, on its child nodes, and so forth. The classes of $L_t$ and $L_{t^c}$ are dependent upon the current node. Step 2 is used to find the optimal $n$ gene pairs and threshold $k$. Step 3 is used to update $X_{t^*}$ into smaller and smaller sample subsets. We iteratively train on the samples remaining in each successive $X_{t^*}$ until gene pairs and thresholds are found for all nodes and leaves of the diagnostic hierarchy.

## c. Identification of the decision-tree marker panel

At each edge of the diagnostic hierarchy, the group of class labels of the parent node is partitioned into class labels of the two children. A classifier is chosen at each edge to direct classification to either one of the child nodes. This classifier is a gene-pair $(g_i, g_j)$ that gives $\Delta_{max}(d_m, d_n)$ between the classes of the two child nodes, $L_m$ and $L_n$. The decision rule for each classifier is: IF $x_{g_i} > x_{g_j}$, THEN classify as phenotype $G_m$; ELSE phenotype $G_n$, where $x_{g_i}$ and $x_{g_j}$ are expression levels of genes $g_i$ and $g_j$, respectively. The collection of gene-pair classifiers at all edges of the diagnostic hierarchy is accumulated vertically into a decision-tree marker panel

(**Table 2**) to guide classification toward a unique leaf in the diagnostic hierarchy. The cumulative binary outcome for an entire route, from root to leaf, delineates the disease-specific molecular signature.

### d. Diagnosis of transcriptome samples

The node marker panel and the diagnostic hierarchy are used for brain phenotype classification. Specifically, the relative expression orderings of the gene-pair classifiers are used to screen for disease-specific expression patterns.

Starting at either child node of the root, we use the corresponding set of $n$ gene pairs and constant threshold $k$. The value of $\mu$ for the transcriptome test sample is compared with $k$. If $\mu \geq k$, the sample is 'positive' for $L_t$ and classification proceeds to both child nodes. However, if $\mu < k$, the sample is 'negative' for $L_t$ and classification stops. We continue this process on the sibling node, on the child nodes, and so forth, unless a sample is deemed 'negative' for a particular node. A test sample can have one of three diagnostic outcomes:

1. A single disease class diagnosis, where the sample is 'positive' for all the nodes of only one entire diagnostic path.

2. No diagnosis, where the sample is 'negative' for at least one node in every path. Here, the sample is rejected from classification, and determined to be none of the brain phenotypes in the diagnostic hierarchy.

3. Multiple diagnoses, where the sample is 'positive' for all the nodes of multiple paths. In this case, the classifiers of the decision-tree marker panel are used as a tie-breaker: All gene-pair classifiers' decision rules are evaluated on the sample, and the resulting binary outcome is compared to the binary signatures of the class candidates. The brain phenotype whose binary signature matches that of the test sample is chosen as the unique diagnosis.

# Text S2. Advantages of using relative expression reversals to build classifiers

Recently, relative expression orderings (i.e. ranks) among a small number of genes have been used for the supervised classification of disease-associated phenotypes [1-3]. The Top-Scoring Pair (TSP) classifier, which uses a single gene-pair within an expression profile that exhibits a characteristic "relative expression reversal" between two classes of interest, has been shown to be a highly accurate and robust strategy in the molecular diagnostics of a variety of human cancers [4-6]. The power of this two-gene classifier method extends beyond high classification performance; by comparing the ranked expression values of only two genes, this simple classification decision rule has remarkably straightforward biological interpretation and eliminates the need for parameter estimation and data normalization. The elimination of the need for data normalization across microarrays is advantageous in the clinic, where presumably the microarray would not be the measurement platform, because the resulting ISSAC based tests require only the genes in the classifier to be measured and do not need all the other genes to assure that a normalization is done uniformly. These are important advantages, since building complex decision rules entails the risk of over-fitting, and data preprocessing may introduce unwanted bias into the dataset. For multi-category diagnoses, we extend on the basic idea of the TSP classifier into a multi-class, coarse-to-fine search strategy [7,8].

## Text S3. Twenty-five anatomical regions of the human brain from which normal transcriptome samples were obtained

Roth *et al.* [1] and Roth *et al.* (GSE7307, unpublished): Accumbens, amygdala, brain stem, caudate, cerebellum, cerebral cortex, entorhinal cortex, frontal lobe, hippocampus, hypothalamus, lateral ventricle, medulla, middle temporal gyrus, occipital lobe, optic nerve, posterior cingulate cortex, prefrontal cortex, posterior fossa, parietal lobe, putamen, superior frontal gyrus, substantia nigra, thalamus, temporal lobe, and visual cortex.

# Text S4. Global statistical enrichment analysis of gene-pair classifiers

Our discussion in the main text on selected marker-panel genes provides some insight into their role in disease. Next, rather than focusing on a limited number of genes (such as only those in our marker-panel), we extended our analysis to large sets of gene pairs that distinguish GBM from OLG. We hypothesized that utilizing more complete information would offer more direct insight into the basis of the classifiers' relative expression reversal behavior, by being able to associate their global patterns with differences between the pathophysiology underlying the two brain cancers.

For any two classes and any given set of gene pairs, the union of the genes that are expressed relatively higher (respectively, lower) in each gene pair for Class 1 will be referred to as 'gene-set $i$' (resp., 'gene-set $j$'). Thus if there are N gene pairs, then 'gene-set $i$' and 'gene-set $j$' each consist of N genes. For the top 500, 1,000, and 1,500 gene pairs (numbers arbitrarily chosen to identify major trends) that best distinguish GBM (Class 1) from OLG (Class 2), we performed an enrichment analysis on the biological process ontologies of 'gene-set $i$' (expressed relatively higher in GBM than in OLG) and on those of 'gene-set $j$' (expressed relatively lower in GBM than in OLG) to find the most consistently enriched category by $Z$-score.

Information on the biological processes and chromosome numbers of genes is available in the PANTHER database [1]. The $Z$-score is defined as:

$$Z_A = \sqrt{n} \, \frac{(P_A - \bar{P}_A)}{\sqrt{\bar{P}_A (1 - \bar{P}_A)}}$$

where $n$ is the total number of classifier gene pairs between two classes (e.g. 500, 1000, or 1,500) and $P_A$ and $\bar{P}_A$ are the proportion of genes characterized by biological category (or chromosome number) $A$ in a given gene set (e.g. gene-set $i$ ) and in the null distribution (i.e. all genes in PANTHER), respectively.

Among 18 major biological processes in the PANTHER database, 'Immunity and Defense' was the most strongly enriched biological process in gene-set $i$ (Figure S4a). Strong enrichment in 'Immunity and Defense' for the genes expressed relatively higher in GBM reflects the frequently observed presence of chronic inflammation in highly malignant cancers [2], such as in GBM [3]. In a tumor-associated inflammatory micro-environment, immune cells penetrate inside the tumor and secrete reactive oxygen species. This can cause further oxidative DNA damage and oncogenic mutations, including amplification of oncogenes or deletion of cell-cycle regulators, and thereby facilitate cancer progression, survival, and migration. Our results show that a relatively highly inflamed tumor environment, composed of a deep infiltration of immune cells and tumor cells exhibiting functions to embattle such oxidative conditions, is the most representative pathophysiological trait that differentiates the tumors of GBM and OLG.

'Neuronal Activities' was the most enriched biological process in gene-set $j$, or the group of genes that are expressed lower in GBM compared to OLG (Figure S4b). This functional category includes basic activities of the nerve or neuron behavior, such as synaptic transmission, neurotransmitter release, and action potential propagation. It has been shown that GBM cells release glutamate, an amino-acid neurotransmitter [4,5]. Elevated levels of extracellular glutamate concentrations is followed by an acute degeneration and death of neurons [4-6], a process known as excitotoxicity, and is one of the underlying causes of tumor-associated epileptic seizures and neuro-cognitive deficiencies in glioblastoma patients [7]. Therefore, this glutamate excitotoxicity in GBM can be the cause of lower normal synaptic transmission and neural function relative to OLG, as is suggested by our enrichment results for gene-set $j$. Glutamate neurotoxicity has also been implicated in other neurodegenerative diseases, including stroke and Alzheimer's disease [8].

Applying the same enrichment analysis strategy described above for chromosome number, we looked for associations between our expression data and gene copy-number alterations frequently observed in GBM and OLG. The genes in gene-set $i$ and gene-set $j$ were the most enriched in Chromosome 1 (Figure S4c) and Chromosome 10 (Figure S4d), respectively. The loss of Chromosome 10 is one of the most frequent genetic aberrations in GBM [9,10], causing the expression of its genes to be heavily suppressed. This offers a possible explanation for the over-representation of Chromosome 10 genes in the gene-set that is expressed relatively less in GBM ('gene-set $j$'), and thereby higher in OLG. The deletion of the short-arm of Chromosome 1 is a hallmark feature of OLG [11,12], which is what we suspect to have caused the over-representation of Chromosome 1 genes in the set that is expressed relatively less in OLG ('gene-set $i$'), and thereby higher in GBM.

The results from our global enrichment analysis in biological processes and chromosome numbers display the relative differences between the collective properties of the two diseases. This offers a holistic view of the major trends that underlie the classifiers' relative expression reversal behavior, which could not have been detected by studying the gene pairs in our marker-panel alone. It is worth noting that we did not observe these same enrichment properties in the GBM-node or OLG-node classifiers in Table 2. This reflects a clear limit to the extent of which disease properties can be explained by using only the minimal number of classifier genes, since those gene pairs were chosen only in the interest of selecting the smallest set with the highest predictive accuracy, regardless of biological relevance.

# Text S5. Whether ISSAC can play a role in identifying misdiagnoses

To know whether a misclassified sample (by ISSAC) had actually been mislabeled by the histopathologist (certainly a distinct possibility) would be a difficult task for two reasons: 1) We are not able to go back to the "questionable" tumor sample and get a re-evaluation of the clinical phenotype as we were not involved in this step; 2) Our markers were trained on all samples labeled as a particular phenotype without a sample filtering stage in our pipeline, thereby taking into consideration all possible sources of heterogeneity (including potentially mislabeled samples). We did this on purpose to not allow ourselves to reject any variance from the published studies, wanting to avoid over-fitting and overly generous accuracy estimations. These conditions make it truly difficult to determine whether a misclassified sample is due to a mislabeling by the histopathologist, or a combination of the above.

# Text S6. Reasoning for selecting only Affymetrix microarray platforms and for not using probe-specific offsets

We chose to not include samples from various platforms and technologies (e.g. other Affymetrix models, Agilent microarrays, RNA-seq) because of their inherent differences in sample preparation steps, hybridization chemistry, probeset/primer length and sequences, data pre-processing techniques, and so forth – all of which lead to poor correlation of the same features, making them not as readily comparable.  This was a choice we made at the beginning of our study, as we wanted to study aspects of reproducibility not associated with platform.  We do acknowledge that using Affymetrix-based datasets (specifically U133A and U133Plus2.0 array chips) is a somewhat restricted case, and that our markers for brain cancers are indeed platform specific (which we state in the manuscript).  However, these platforms are still the most widely represented in GEO for brain cancers (over 50%).  In addition, we do not anticipate a significant change in the main message or impact of our manuscript if we were to apply our technique to a smaller sample collection of other platforms and technologies, although, of course, the classifier features themselves will likely change.

To the best of our knowledge, there are no confirmed/validated mapping steps (or offsets) that one would routinely apply on a subset of probesets for the cross-sample or cross-platform purposes.  Furthermore, a simple comparison (such as relative expression comparison used throughout this study) is invariant to any monotone pre-processing (in particular any linear transformation) of the raw expression values.  However, with an offset, the outcome of the pair comparison would no longer be invariant to even scaling.  Thus one would lose robustness.  One of the motivations for our method, and one of the reasons relative expression comparison has worked in the past, is this invariance.

# Text S7. Candidates of brain cancer molecular signatures

We would like to provide a cautionary point that more work is necessary prior to any application of our results to a clinical setting. Indeed, we acknowledge that our markers can only serve as classifier candidates at this time. While we were eager to confirm our markers, we did not have brain cancer biopsies at the beginning of our study to validate by qPCR. Nevertheless, we feel that our brain cancer marker candidates, and methods used for their discovery presented herein, are valuable resources to those interested in identifying novel molecular signatures from high-throughput biomolecular data.

# Supplementary Figures and Tables

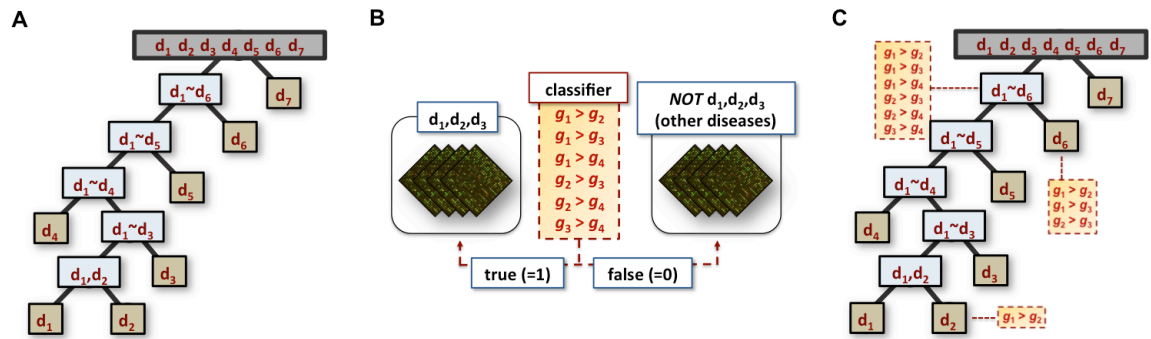

**Figure S1.** The overall method of ISSAC can be summarized into three main steps. **A** ISSAC constructs the framework for brain cancer diagnosis – a tree-structured hierarchy of all brain cancer phenotypes built using an agglomerative hierarchical clustering algorithm on gene expression training data. **B** Training on gene-expression data from all brain phenotypes, ISSAC identifies disjoint, gene-pair classifiers at all nodes (excluding the root) and edges of the diagnostic hierarchy, and accumulates them into their respective marker panels. The chosen pairs are the ones that best differentiate between the phenotype sets, and are based entirely on the *reversal of relative expression*. **C** ISSAC uses the gene-pair classifiers for class prediction. Briefly, given a gene expression profile, ISSAC executes the node classifiers in a hierarchical, top-down fashion within the disease diagnostic hierarchy to identify the phenotype(s) whose class-specific signature(s) is present. In case of multiple class candidates (i.e. node classifiers for multiple leaves are positive), the ambiguity is resolved by aggregating all the decision-tree classifiers into a classification decision-tree, thereby leading any expression signature down one unique path toward a single phenotype.

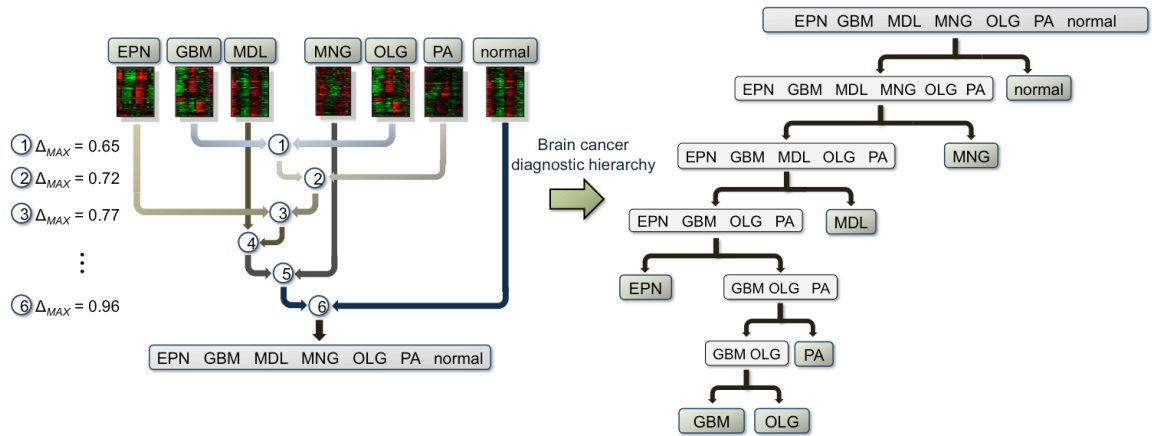

**Figure S2.** Brain phenotypes are grouped into a global diagnostic hierarchy, which allows an intuitive representation of the classification process. The diagnostic hierarchy is built using a data-driven, iterative approach, and is free of manual, ad-hoc construction. In each iteration, two classes, or two groups of classes, with the lowest TSP score (Materials and Methods and Text S1) among all pair-wise comparisons, come together to form a node. This approach optimizes overall classification by placing the more challenging decisions further away from the base of the tree (i.e. root), thereby ensuring only the minimum misclassifications percolate down the tree. The final form of the brain phenotype diagnostic hierarchy represents a hierarchical structure of nested partitions, where the multi-class problem is decomposed into smaller and smaller groups using a sequence of diagnostic decision rules.

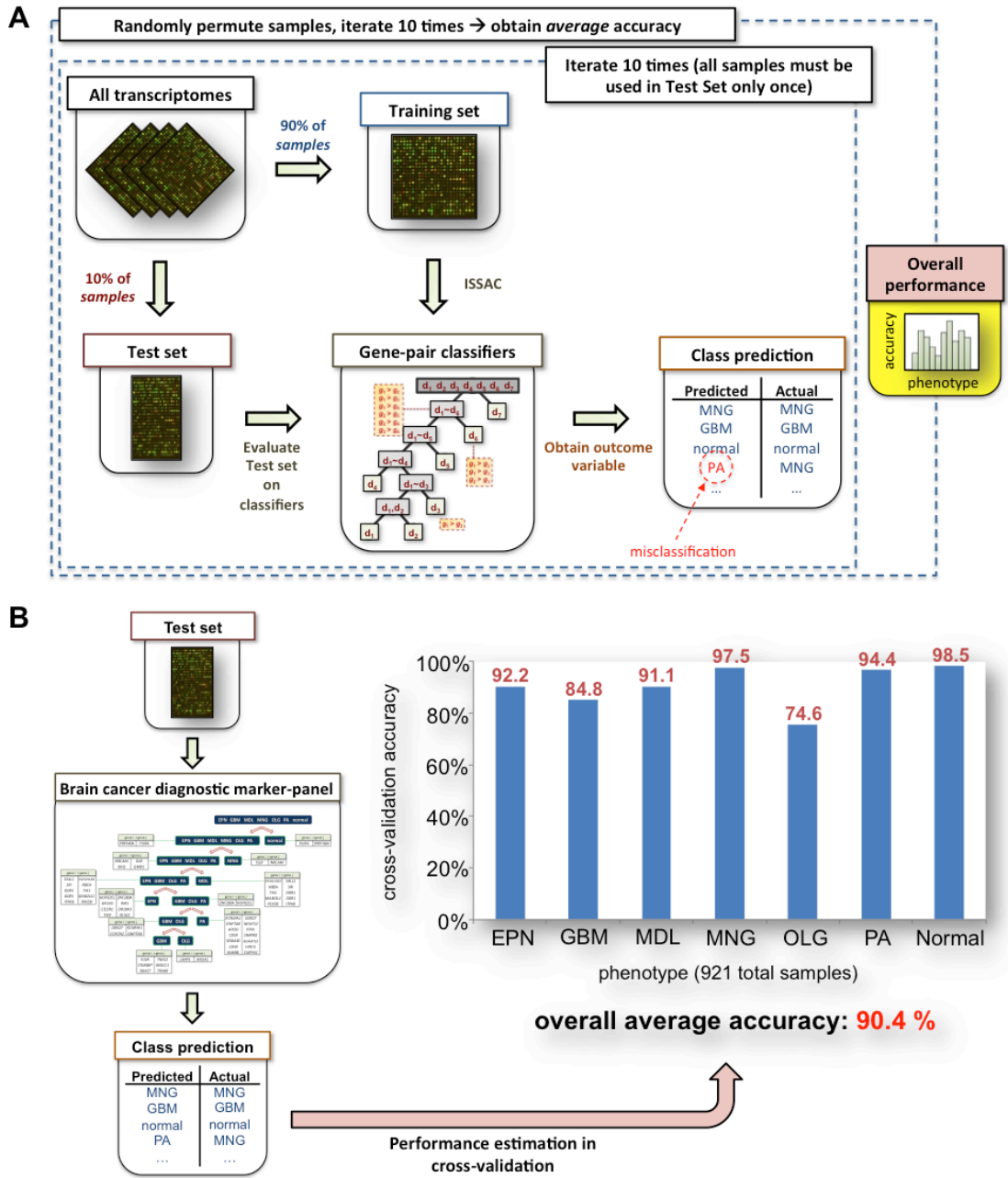

**Figure S3.** Performance evaluation using ten-fold cross-validation. **A** Ten-fold cross-validation is conducted ten times to obtain the average accuracy. In every iteration of cross-validation, the order of samples within a particular class are randomly permuted before training/test set allocations. **B** Our marker panel achieved a 90.4% average of phenotype-specific classification accuracies, showing strong promise against a multi-category, multi-dataset background at the gene expression level.

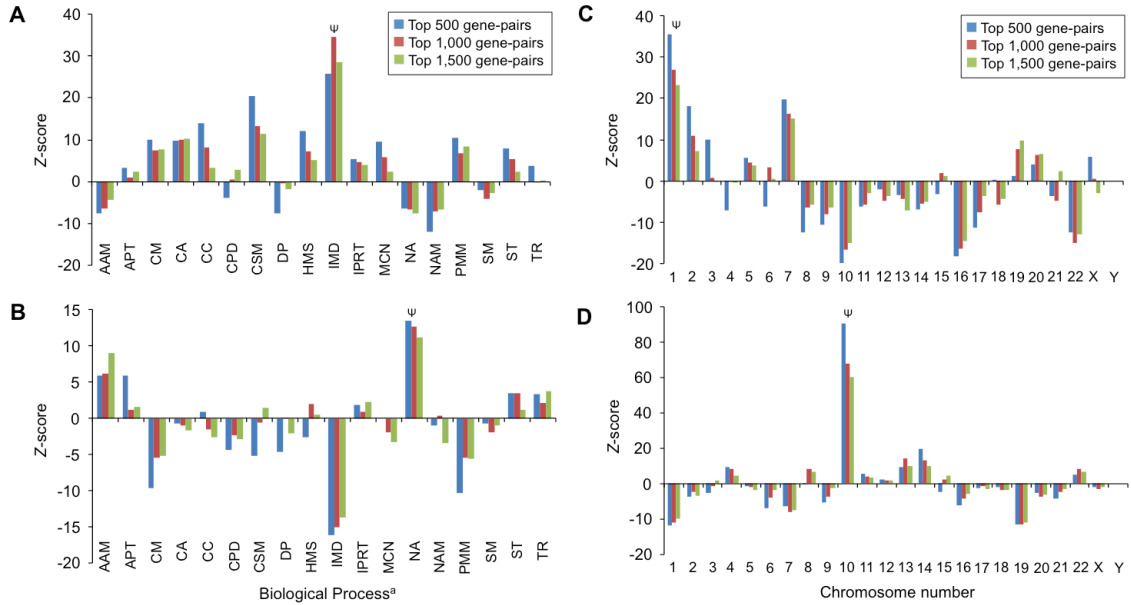

**Figure S4.** Statistical enrichment analysis on PANTHER database biological processes and chromosome numbers of the top 500, 1,000, and 1,500 gene-pair classifiers for GBM vs. OLG. **A** 'Immunity and Defense' was the most enriched biological process for 'gene-set *i*', reflecting the chronic inflammatory conditions inside the GBM tumor. **B** 'Neuronal Activities' was the most enriched biological process for 'gene-set *j*', reflecting decrease in neuronal behavior and possibly other brain cell activity inside the GBM tumor. The genes in 'gene-set *i*' and 'gene-set *j*' were the most enriched in **C** Chromosome 1 and **D** Chromosome 10, respectively, reflecting the major chromosome aberrations of the two brain cancers. Ψ delineates the most enriched category.
[a]Biological process abbreviation (name): AAM (Amino acid metabolism), APT (Apoptosis), CM (Carbohydrate metabolism), CA (Cell adhesion), CC (Cell cycle), CPD (Cell proliferation and differentiation), CSM (Cell structure and motility), DP (Developmental processes), HMS (Homeostatis), IMD (Immunity and defense), IPRT (Intracellular protein transport), MCN (Muscle contraction), NA (Neuronal activities), NAM (Nucleic acid metabolism), PMM (Protein metabolism and modification), SM (Sulfur metabolism), ST (Signal transduction), and TR (Transport).

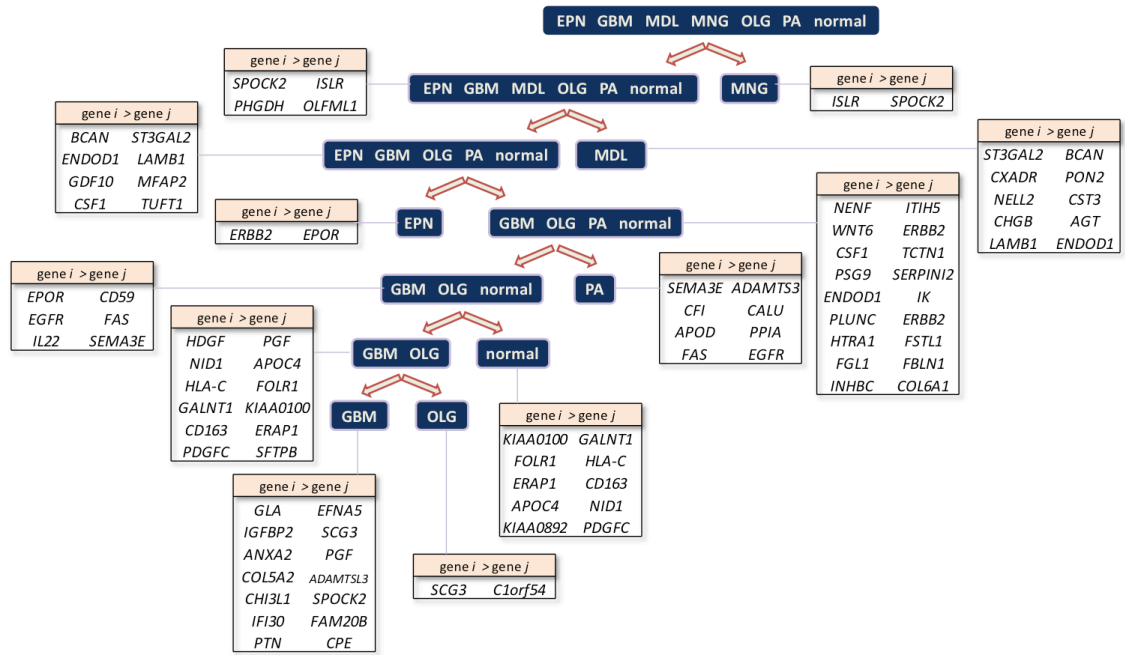

**Figure S5.** Gene-pair classifiers based on only the genes that encode extracellular products. Gene pairs are shown at their corresponding nodes in the brain disease diagnostic hierarchy. The corresponding node-based marker panel consists of 41 classifier pairs and 71 unique classifier features.

**Table S1.** Phenotype specimen descriptions and main results for all GEO accessions used in this study.

| Phenotype Name | GEO accession # | First Author (publication year) | Phenotype specimen description | Main results |
|---|---|---|---|---|
| Ependymoma | GSE16155 | Donson (2009) | • Human ependymoma tumor resections | • Genes associated with nonrecurrent ependymoma were predominantly immune function-related<br>• Histological analysis of a subset of immune function genes revealed that their expression was restricted to tumor-infiltrating subpopulation<br>• Up-regulation of immune function genes is the predominant ontology associated with a good prognosis in ependymoma |
| | GSE21687 | Johnson (2010) | • Human ependymomas comprised of minimum 85% tumour cells | • Identified subgroups of ependymoma, and subgroup-specific gene amplifications and deletions<br>• Comparative transcriptomics between human tumors and mouse neural stem cells generated mouse models of ependymoma with matching molecular expression patterns<br>• Developed a novel cross-species genomic approach to match subgroup-specific driver mutations with cellular compartments to model cancer subgroups |
| Glioblastoma Multiforme | GSE4412 | Freije (2004) | • Diffuse infiltrating gliomas | • Gene expression-based grouping of tumors is a more powerful survival predictor than histologic grade or age<br>• The expression patterns of 44 genes classify gliomas into previously unrecognized biological and prognostic groups<br>• Large-scale gene expression analysis and subset analysis of gliomas reveals unrecognized heterogeneity of tumors |
| | GSE4271 | Phillips (2006) | • Primary high-grade gliomas and matched recurrences | • Novel prognostic subclasses of high-grade astrocytoma closely resemble stages in neurogenesis<br>• One tumor class displaying neuronal lineage markers shows longer survival, while two tumor classes enriched for neural stem cell markers display equally short survival<br>• Poor prognosis subclasses exhibit either markers of proliferation or of angiogenesis and mesenchyme<br>• A robust two-gene prognostic model utilizing PTEN and DLL3 expression suggests that Akt and Notch signaling are hallmarks of poor prognosis versus better prognosis gliomas, respectively |
| | GSE8692 | Liu (2007) | • Primary low/high grade gliomas | • Measured genome-wide mRNA expression levels and miRNA profiles by microarray analysis and RT-PCR, respectively<br>• Correlation coefficients were determined for all possible mRNA-miRNA pairs<br>• A subset of high correlated pairs were experimentally validated by overexpressing or suppressing a miRNA and measuring the correlated mRNAs |
| | GSE9171 | Wiedemeyer (2008) | • Glioblastoma tumors | • A nonheuristic genome topography scan (GTS) algorithm was developed to characterize the patterns of genomic alterations in human glioblastoma (GBM)<br>• A codeletion pattern found among closely related INK genes in the GBM oncogenome challenges the prevailing single-hit model of RB pathway inactivation<br>• Results suggest a feedback regulatory circuit in the astrocytic lineage and demonstrate a bona fide tumor suppressor role for $p18^{INK4c}$ in human GBM |
| | GSE4290 | Sun (2006) | • Primary gliomas and nontumor brain samples | • Stem cell factor (SCF) activates brain microvascular endothelial cells in vitro and induces a potent angiogenic response in vivo<br>• SCF downregulation inhibits tumor-mediated angiogenesis and glioma growth, whereas SCF overexpression is associated with shorter survival in malignant glioma patients<br>• The SCF/c-Kit pathway plays an important role in tumor- and normal host cell-induced angiogenesis within the brain<br>• Anti-angiogenic strategies have great potential as a treatment approach for gliomas |
| Medulloblastoma | GSE10327 | Kool (2008) | • Primary medulloblastomas and local relapses | • mRNA expression profiling and genomic hybridization arrays show 5 different types of medulloblastoma, each with characteristic pathway activation signatures and associated specific genetic defects<br>• Clinicopathological features significantly different between the 5 subtypes include metastatic disease, age at diagnosis, and histology |
| | GSE12992 | Fattet (2009) | • Paediatric medulloblastomas | • Immunostaining of β-catenin showed extensive nuclear staining in a subset of samples<br>• Expression profiles show strong activation of the Wnt/β-catenin pathway, and complete loss of chromosome 6<br>• Patients with extensive nuclear staining were significantly older at diagnosis and were in complete remission after a mean follow-up of 75.7 months (range 27.5–121.2 months) from diagnosis<br>• Results confirm previous observations that CTNNB1-mutated tumours represent a distinct molecular subgroup of medulloblastomas with favourable outcome |

**Table S1.** (Continued) Phenotype specimen descriptions and main results for all GEO accessions used in this study.

| Phenotype Name | GEO accession # | First Author (publication year) | Phenotype specimen description | Main results |
|---|---|---|---|---|
| Meningioma | GSE4780 | Scheck (2006) | • Benign (grade 1) and aggressive (grades 2 and 3) meningiomas | • The results of this study have not been publicly disclosed (only data are available) |
| | GSE9438 | Claus (2008) | • Meningioma specimens without neurofibromatosis type 2, nonrecurrent | • Progesterone and estrogen hormone receptors (PR and ER, respectively) were measured via immunohistochemistry and compared with gene expression profiling results<br>• Gene expression seemed more strongly associated with PR status (+/-) than with ER status<br>• Genes in collagen and extracellular matrix pathways were most differentially expressed by PR status<br>• PR status may be a clinical marker for genetic subgroups of meningioma |
| | GSE 16581 | Lee (2010) | • 85 meningioma samples of various grades | • Investigators performed retrospective global genetic and genomic analysis of 85 meningioma samples of various grades<br>• In addition to chromosome 22q loss, which was detected in the majority of clinical samples, chromosome 6q and 14q loss was significantly more common in recurrent tumors and was associated with anaplastic histology<br>• Five "classes" of meningiomas were detected by gene expression analysis that correlated with copy number alterations, recurrent status, and malignant histology<br>• Data from this study provide broad genomic information to further stratify meningioma patients into prognostic risk groups |
| Oligodendroglioma | GSE4412 | Freije (2004) | • Primary high-grade gliomas and matched recurrences | • Novel prognostic subclasses of high-grade astrocytoma are identified and discovered to resemble stages in neurogenesis<br>• One tumor class displaying neuronal lineage markers shows longer survival, while two tumor classes enriched for neural stem cell markers display equally short survival<br>• Poor prognosis subclasses exhibit either markers of proliferation or of angiogenesis and mesenchyme<br>• A robust two-gene prognostic model utilizing PTEN and DLL3 expression suggests that Akt and Notch signaling are hallmarks of poor prognosis versus better prognosis gliomas, respectively |
| | GSE4290 | Sun (2006) | • Primary gliomas and nontumor brain samples | • Stem cell factor (SCF) activates brain microvascular endothelial cells in vitro and induces a potent angiogenic response in vivo<br>• Downregulation of SCF inhibits tumor-mediated angiogenesis and glioma growth in vivo, whereas overexpression of SCF is associated with shorter survival in patients with malignant gliomas<br>• The SCF-c-Kit pathway plays an important role in tumor- and normal host cell-induced angiogenesis within the brain<br>• Antiangiogenic strategies have great potential as a treatment approach for gliomas |
| Pilocytic Astrocytoma | GSE 12907 | Wong (2005) | • Juvenile pilocytic astrocytomas (JPAs) | • Genes involved in certain biological processes, including neurogenesis, cell adhesion, and central nervous system development, were significantly deregulated in JPA compared to those in normal cerebella<br>• Two major subgroups of JPA based on unsupervised hierarchical clustering<br>• JPA without myelin basic protein-positively stained tumor cells may have a higher tendency to progress |
| | GSE 5675 | Sharma (2007) | • Primary pilocytic astrocytomas (PAs) arising sporadically and in patients with neurofibromatosis type 1 (NF1) | • No expression signature to discriminate clinically aggressive/recurrent tumors from indolent<br>• Unique gene expression pattern for PAs arising in patients with NF1<br>• Gene expression signature stratified PAs by location (supratentorial versus infratentorial)<br>• Glial tumors may share an intrinsic, lineage-specific molecular signature that reflects the brain region in which their nonmalignant predecessors originated |
| Normal Brain | GSE3526 | Roth (2006) | • 20 anatomically distinct sites of the central nervous system (CNS)<br>• 8 autopsies for each CNS region<br>• Patient death was due to sudden death | • Principal component analysis and hierarchical clustering results showed that the expression patterns of the 20 CNS sites profiled were significantly different from all non-CNS tissues and were also similar to one another, indicating an underlying common expression signature<br>• The 20 sites could be segregated into discrete groups with underlying similarities in anatomical structure and, in many cases, functional activity |
| | GSE7307 | Roth (2007) | • Normal and diseased human tissues representing over 90 distinct tissue types<br>• Patient death was due to sudden death | • The results of this study have not been publicly disclosed (only data are available) |

**Supplementary Table 2.** GEO microarray sample IDs used in this study.

| Phenotype Name | Phenotype Label | GEO Accession Series # | GEO Microarray Sample ID |
|---|---|---|---|
| Ependymoma | EPN | GSE16155 | GSM404936,GSM404937,GSM404938,GSM404939,GSM404940,GSM404941,GSM404942,GSM404943,GSM404944,GSM404945,GSM404946,GSM404947,GSM404948,GSM404949,GSM404950,GSM404951,GSM404952,GSM404953,GSM404954 |
| | | GSE21687 | GSM541060,GSM541061,GSM541062,GSM541063,GSM541064,GSM541065,GSM541066,GSM541067,GSM541068,GSM541069,GSM541070,GSM541071,GSM541072,GSM541073,GSM541074,GSM541075,GSM541076,GSM541077,GSM541078,GSM541079,GSM541080,GSM541081,GSM541082,GSM541083,GSM541084,GSM541085,GSM541086,GSM541087,GSM541088,GSM541089,GSM541090,GSM541091,GSM541092,GSM541093,GSM541094,GSM541095,GSM541096,GSM541097,GSM541098,GSM541099,GSM541100,GSM541101,GSM541102,GSM541103,GSM541104,GSM541105,GSM541106,GSM541107,GSM541108,GSM541109,GSM541110,GSM541111,GSM541112,GSM541113,GSM541114,GSM541115,GSM541116,GSM541117,GSM541118,GSM541119,GSM541120,GSM541121,GSM541122,GSM541123,GSM541124,GSM541125,GSM541126,GSM541127,GSM541128,GSM541129,GSM541130,GSM541131,GSM541132,GSM541133,GSM541134,GSM541135,GSM541136,GSM541137,GSM541138,GSM541139,GSM541140,GSM541141,GSM541142 |
| Glioblastoma Multiforme | GBM | GSE 4412 | GSM99432,GSM99434,GSM99436,GSM99438,GSM99440,GSM99442,GSM99444,GSM99446,GSM99448,GSM99450,GSM99452,GSM99454,GSM99456,GSM99462,GSM99464,GSM99466,GSM99470,GSM99472,GSM99474,GSM99476,GSM99478,GSM99480,GSM99482,GSM99484,GSM99486,GSM99488,GSM99490,GSM99492,GSM99494,GSM99524,GSM99526,GSM99528,GSM99530,GSM99532,GSM99534,GSM99536,GSM99538,GSM99540,GSM99542,GSM99544,GSM99546,GSM99548,GSM99550,GSM99552,GSM99554,GSM99556,GSM99558,GSM99560,GSM99562,GSM99564,GSM99572,GSM99576,GSM99578,GSM99580,GSM99582,GSM99584,GSM99586,GSM99588,GSM99590 |
| | | GSE 4271 | GSM96950,GSM96951,GSM96952,GSM96953,GSM96954,GSM96955,GSM96956,GSM96957,GSM96958,GSM96959,GSM96960,GSM96961,GSM96962,GSM96963,GSM96964,GSM96965,GSM96966,GSM96967,GSM96968,GSM96969,GSM96970,GSM96971,GSM96972,GSM96973,GSM96974,GSM96975,GSM96976,GSM96977,GSM96978,GSM96979,GSM96980,GSM96981,GSM96982,GSM96983,GSM96984,GSM96985,GSM96986,GSM96987,GSM96988,GSM96989,GSM96990,GSM96991,GSM96992,GSM96993,GSM96994,GSM96995,GSM96996,GSM96997,GSM96998,GSM96999,GSM97000,GSM97001,GSM97002,GSM97003,GSM97004,GSM97005,GSM97006,GSM97007,GSM97008,GSM97009,GSM97010,GSM97011,GSM97014,GSM97018,GSM97021,GSM97024,GSM97028,GSM97031,GSM97032,GSM97037,GSM97040,GSM97041,GSM97042,GSM97044,GSM97048,GSM97049 |
| | | GSE 8692 | GSM215420,GSM215422,GSM215423,GSM215425,GSM215426,GSM215427 |
| | | GSE 9171 | GSM231695,GSM231696,GSM231697,GSM231698,GSM231699,GSM231700,GSM231701,GSM231702,GSM231703,GSM231704,GSM231705,GSM231706,GSM231707 |
| | | GSE 4290 | GSM97794,GSM97796,GSM97797,GSM97798,GSM97801,GSM97806,GSM97808,GSM97813,GSM97814,GSM97818,GSM97819,GSM97821,GSM97829,GSM97832,GSM97839,GSM97844,GSM97847,GSM97851,GSM97852,GSM97856,GSM97859,GSM97861,GSM97863,GSM97869,GSM97870,GSM97871,GSM97877,GSM97882,GSM97885,GSM97886,GSM97887,GSM97888,GSM97889,GSM97891,GSM97892,GSM97893,GSM97894,GSM97895,GSM97896,GSM97898,GSM97903,GSM97905,GSM97906,GSM97908,GSM97912,GSM97914,GSM97915,GSM97917,GSM97918,GSM97919,GSM97922,GSM97924,GSM97926,GSM97930,GSM97931,GSM97935,GSM97936,GSM97938,GSM97940,GSM97942,GSM97945,GSM97946,GSM97948,GSM97950,GSM97952,GSM97953,GSM97954,GSM97955,GSM97959,GSM97961,GSM97963,GSM97965,GSM97966,GSM97967,GSM97968,GSM97969,GSM97971 |
| Medulloblastoma | MDL | GSE 10327 | GSM260959,GSM260960,GSM260961,GSM260962,GSM260963,GSM260964,GSM260965,GSM260966,GSM260967,GSM260968,GSM260969,GSM260970,GSM260971,GSM260972,GSM260973,GSM260974,GSM260975,GSM260976,GSM260977,GSM260978,GSM260979,GSM260980,GSM260981,GSM260982,GSM260983,GSM260984,GSM260985,GSM260986,GSM260987,GSM260988,GSM260989,GSM260990,GSM260991,GSM260992,GSM260993,GSM260994,GSM260995,GSM260996,GSM260997,GSM260998,GSM260999,GSM261000,GSM261001,GSM261002,GSM261003,GSM261004,GSM261005,GSM261006,GSM261007,GSM261008,GSM261009,GSM261010,GSM261011,GSM261012,GSM261013,GSM261014,GSM261015,GSM261016,GSM261017,GSM261018,GSM261019,GSM261020 |
| | | GSE 12992 | GSM324062,GSM324063,GSM324064,GSM324065,GSM324066,GSM324067,GSM324068,GSM324069,GSM324082,GSM324083,GSM324084,GSM324085,GSM324090,GSM324091,GSM324092,GSM324093,GSM324104,GSM324111,GSM324112,GSM324113,GSM324115,GSM324119,GSM324137,GSM324138,GSM324139,GSM324140,GSM324141,GSM324508,GSM324512,GSM324513,GSM324514,GSM324515,GSM324516,GSM324517,GSM324526,GSM325233,GSM325278,GSM325280,GSM325281,GSM325282 |
| Meningioma | MNG | GSE 4780 | GSM108014,GSM108015,GSM108016,GSM108017,GSM108018,GSM107987,GSM107988,GSM107989,GSM107990,GSM107991,GSM107992,GSM107993,GSM107994,GSM107995,GSM107996,GSM107997,GSM107998,GSM107999,GSM108000,GSM108001,GSM108002,GSM108003,GSM108004,GSM108005,GSM108006,GSM108007,GSM108008,GSM108009,GSM108010,GSM108011,GSM108012,GSM108013,GSM108019,GSM108020,GSM108021,GSM108022,GSM108023,GSM108024,GSM108025,GSM108026,GSM108027,GSM108028,GSM108029,GSM108030,GSM108031,GSM108032,GSM108033,GSM108034,GSM108035,GSM108036,GSM108037,GSM108038,GSM108039,GSM108040,GSM108041,GSM108042,GSM108043,GSM108044,GSM108045,GSM108046,GSM108047,GSM108048 |
| | | GSE 9438 | GSM239770,GSM239771,GSM239772,GSM239773,GSM239774,GSM239775,GSM239776,GSM239777,GSM239778,GSM239779,GSM239780,GSM239781,GSM239782,GSM239783,GSM239784,GSM239785,GSM239786,GSM239787,GSM239788,GSM239789,GSM239790,GSM239791,GSM239792,GSM239793,GSM239794,GSM239795,GSM239796,GSM239797,GSM239798,GSM239799,GSM239800 |
| | | GSE 16581 | GSM416798,GSM416799,GSM416800,GSM416801,GSM416802,GSM416803,GSM416804,GSM416805,GSM416806,GSM416807,GSM416808,GSM416809,GSM416810,GSM416811,GSM416812,GSM416813,GSM416814,GSM416815,GSM416816,GSM416817,GSM416818,GSM416819,GSM416820,GSM416821,GSM416822,GSM416823,GSM416824,GSM416825,GSM416826,GSM416827,GSM416828,GSM416829,GSM416830,GSM416831,GSM416832,GSM416833,GSM416834,GSM416835,GSM416836,GSM416837,GSM416838,GSM416839,GSM416840,GSM416841,GSM416842,GSM416843,GSM416844,GSM416845,GSM416846,GSM416847,GSM416848,GSM416849,GSM416850,GSM416851,GSM416852,GSM416853,GSM416854,GSM416855,GSM416856,GSM416857,GSM416858,GSM416859,GSM416860,GSM416861,GSM416862,GSM416863,GSM416864,GSM416865 |
| Oligodendroglioma | OLG | GSE 4412 | GSM99458,GSM99460,GSM99510,GSM99512,GSM99514,GSM99516,GSM99518,GSM99520,GSM99522,GSM99570,GSM99598 |
| | | GSE 4290 | GSM97799,GSM97822,GSM97823,GSM97824,GSM97830,GSM97831,GSM97835,GSM97838,GSM97841,GSM97842,GSM97845,GSM97854,GSM97857,GSM97860,GSM97862,GSM97864,GSM97865,GSM97866,GSM97867,GSM97868,GSM97872,GSM97873,GSM97874,GSM97875,GSM97876,GSM97880,GSM97881,GSM97883,GSM97884,GSM97897,GSM97900,GSM97901,GSM97902,GSM97904,GSM97907,GSM97909,GSM97911,GSM97923,GSM97925,GSM97928,GSM97929,GSM97933,GSM97934,GSM97944,GSM97947,GSM97949,GSM97956,GSM97962,GSM97964,GSM97970 |

**Supplementary Table 2.** (Continued) GEO microarray sample IDs used in this study.

| | | | |
|---|---|---|---|
| Pilocytic Astrocytoma | PA | GSE 12907 | GSM322969,GSM323054,GSM323523,GSM323524,GSM323525,GSM323526,GSM323527,GSM323528,GSM323529,GSM323530,GSM323531,GSM323554,GSM323555,GSM323557,GSM323558,GSM323559,GSM323560,GSM323561,GSM323562,GSM323563,GSM323564 |
| | | GSE 5675 | GSM132714,GSM132715,GSM132716,GSM132717,GSM132718,GSM132719,GSM132720,GSM132722,GSM132723,GSM132728,GSM132729,GSM132730 ,GSM132733,GSM132736,GSM132738 ,GSM132741 ,GSM132744,GSM132747,GSM132748,GSM132750 ,GSM132751,GSM132752,GSM132753,GSM132754,GSM132759,GSM132761,GSM132763,GSM132765,GSM132768,GSM132769,GSM132770,GSM132771,GSM132772,GSM132773,GSM132774,GSM132775,GSM132776,GSM132777,GSM132778,GSM132779,GSM132780 |
| Normal Brain | normal | GSE 3526 | GSM80565,GSM80566,GSM80567,GSM80568,GSM80569,GSM80570,GSM80571,GSM80572,GSM80573,GSM80574,GSM80575,GSM80581,GSM80585,GSM80586,GSM80587,GSM80591,GSM80592,GSM80593,GSM80594,GSM80595,GSM80596,GSM80597,GSM80598,GSM80599,GSM80600,GSM80601,GSM80616,GSM80617,GSM80618,GSM80619,GSM80620,GSM80621,GSM80622,GSM80623,GSM80626,GSM80627,GSM80628,GSM80636,GSM80637,GSM80638,GSM80639,GSM80640,GSM80641,GSM80642,GSM80643,GSM80644,GSM80645,GSM80646,GSM80647,GSM80650,GSM80651,GSM80652,GSM80653,GSM80660,GSM80661,GSM80662,GSM80663,GSM80664,GSM80665,GSM80666,GSM80667,GSM80668,GSM80669,GSM80670,GSM80671,GSM80675,GSM80676,GSM80677,GSM80678,GSM80679,GSM80680,GSM80681,GSM80682,GSM80683,GSM80684,GSM80690,GSM80691,GSM80692,GSM80693,GSM80699,GSM80700,GSM80701,GSM80702,GSM80703,GSM80704,GSM80705,GSM80706,GSM80708,GSM80709,GSM80711,GSM80713,GSM80714,GSM80715,GSM80721,GSM80722,GSM80723,GSM80724,GSM80744,GSM80745,GSM80746,GSM80747,GSM80752,GSM80754,GSM80756,GSM80760,GSM80761,GSM80762,GSM80763,GSM80766,GSM80767,GSM80772,GSM80773,GSM80774,GSM80775,GSM80800,GSM80801,GSM80802,GSM80803,GSM80804,GSM80817,GSM80818,GSM80819,GSM80830,GSM80831,GSM80832,GSM80833,GSM80834,GSM80835,GSM80836,GSM80837,GSM80838,GSM80839,GSM80840,GSM80841,GSM80847,GSM80848,GSM80849,GSM80851,GSM80852,GSM80855,GSM80858,GSM80859,GSM80860,GSM80861,GSM80862,GSM80863 |
| | | GSE 7307 | GSM175842,GSM175843,GSM175844,GSM175845,GSM175849,GSM175850,GSM175851,GSM176153,GSM175852,GSM175853,GSM175854,GSM176048,GSM175855,GSM175856,GSM175857,GSM175858,GSM175846,GSM175847,GSM175848,GSM176150,GSM175871,GSM175872,GSM175873,GSM176059,GSM175874,GSM175875,GSM175876,GSM175877,GSM176215,GSM175901,GSM175902,GSM175903,GSM175904,GSM175987,GSM175988,GSM176174,GSM176175,GSM175989,GSM175990,GSM176170,GSM176171,GSM176172,GSM176173,GSM176178,GSM176179,GSM176180,GSM176181,GSM176182,GSM176183,GSM176184,GSM176185,GSM176161,GSM176162,GSM176163,GSM176164,GSM176056,GSM176073 |

**Table S3.** Node marker panel for brain cancer and normal transcriptome classification.

| Node # | Node phenotype classes | Gene i | | | | Gene j | | | | k |
|---|---|---|---|---|---|---|---|---|---|---|
| | | Gene symbol | Gene name | Chromosome locus | Affymetrix Probe ID | Gene symbol | Gene name | Chromosome locus | Affymetrix Probe ID | |
| 2 | EPN GBM MDL MNG OLG PA | PRPF40A | PRP40 pre-mRNA processing factor 40 homolog A (S. cerevisiae) | 2q23.3 | 218053_at | PURA | Purine-rich element binding protein A | 5q31 | 204021_s_at | 1 |
| 3 | normal | PURA | Purine-rich element binding protein A | 5q31 | 204021_s_at | PRPF40A | PRP40 pre-mRNA processing factor 40 homolog A (S. cerevisiae) | 2q23.3 | 218053_at | 1 |
| 4 | EPN GBM MDL OLG PA | NRCAM | Neuronal cell adhesion molecule | 7q31 | 204105_s_at | ISLR | Immunoglobulin superfamily containing leucine-rich repeat | 15q23-q24 | 207191_s_at | 1 |
| | | IDH2 | Isocitrate dehydrogenase 2 (NADP+), mitochondrial | 15q26.1 | 210046_s_at | GMDS | GDP-mannose 4,6-dehydratase | 6p25 | 214106_s_at | |
| 5 | MNG | ISLR | Immunoglobulin superfamily containing leucine-rich repeat | 15q23-q24 | 207191_s_at | NRCAM | Neuronal cell adhesion molecule | 7q31 | 204105_s_at | 1 |
| 6 | EPN GBM OLG PA | SALL1 | Sal-like 1 (Drosophila) | 16q12.1 | 206893_at | PAFAH1B3 | Platelet-activating factor acetylhydrolase 1b, catalytic subunit 3 | 19q13.1 | 203228_at | 2 |
| | | SRI | Sorcin | 7q21 | 208920_at | NBEA | Neurobeachin | 13q13 | 221207_s_at | |
| | | DDR1 | Discoidin domain receptor tyrosine kinase 1 | 6p21.3 | 210749_x_at | TIA1 | TIA1 cytotoxic granule-associated RNA binding protein | 2p13 | 201447_at | |
| | | DDR1 | Discoidin domain receptor tyrosine kinase 1 | 6p21.3 | 208779_x_at | MAB21L1 | Mab-21-like 1 (C. elegans) | 13q13 | 206163_at | |
| | | ITPKB | Inositol 1,4,5-trisphosphate 3-kinase B | 1q42.13 | 203723_at | PDS5B | PDS5, regulator of cohesion maintenance, homolog B (S. cerevisiae) | 13q12.3 | 204742_s_at | |
| 7 | MDL | PAFAH1B3 | Platelet-activating factor acetylhydrolase 1b, catalytic subunit 3 | 19q13.1 | 203228_at | SALL1 | Sal-like 1 (Drosophila) | 16q12.1 | 206893_at | 4 |
| | | NBEA | Neurobeachin | 13q13 | 221207_s_at | SRI | Sorcin | 7q21 | 208920_at | |
| | | TIA1 | TIA1 cytotoxic granule-associated RNA binding protein | 2p13 | 201447_at | DDR1 | Discoidin domain receptor tyrosine kinase 1 | 6p21.3 | 210749_x_at | |
| | | MAB21L1 | Mab-21-like 1 (C. elegans) | 13q13 | 206163_at | DDR1 | Discoidin domain receptor tyrosine kinase 1 | 6p21.3 | 208779_x_at | |
| | | PDS5B | PDS5, regulator of cohesion maintenance, homolog B (S. cerevisiae) | 13q12.3 | 204742_s_at | ITPKB | Inositol 1,4,5-trisphosphate 3-kinase B | 1q42.13 | 203723_at | |
| 8 | EPN | NUP62CL | Nucleoporin 62kDa C-terminal like | Xq22.3 | 220520_s_at | ZNF280A | Zinc finger protein 280A | 22q11.22 | 216034_at | 2 |
| | | GALNS | Galactosamine (N-acetyl)-6-sulfate sulfatase | 16q24.3 | 206335_at | WAS | Wiskott-Aldrich syndrome (eczema-thrombocytopenia) | Xp11.4-p11.21 | 38964_r_at | |
| | | CELSR1 | Cadherin, EGF LAG seven-pass G-type receptor 1 (flamingo homolog, Drosophila) | 22q13.3 | 41660_at | OR10H3 | Olfactory receptor, family 10, subfamily H, member 3 | 19p13.1 | 208520_at | |
| | | TLE4 | Transducin-like enhancer of split 4 (E(sp1) homolog, Drosophila) | 9q21.31 | 216997_x_at | OLIG2 | Oligodendrocyte lineage transcription factor 2 | 21q22.11 | 213824_at | |

| Node # | Node phenotype classes | Gene i | | | | Gene j | | | | k |
|---|---|---|---|---|---|---|---|---|---|---|
| | | Gene symbol | Gene name | Chromosome locus | Affymetrix Probe ID | Gene symbol | Gene name | Chromosome locus | Affymetrix Probe ID | |
| 9 | GBM OLG PA | ZNF280A | Zinc finger protein 280A | 22q11.22 | 216034_at | NUP62CL | Nucleoporin 62kDa C-terminal like | Xq22.3 | 220520_s_at | 1 |
| 10 | GBM OLG | DDX27 | DEAD (Asp-Glu-Ala-Asp) box polypeptide 27 | 20q13.13 | 215693_x_at | KCNMA1 | Potassium large conductance calcium-activated channel, subfamily M, alpha member 1 | 10q22.3 | 221584_s_at | 1 |
| | | COX7A2 | Cytochrome c oxidase subunit VIIa polypeptide 2 (liver) | 6q12 | 217249_at | GNPTAB | N-acetylglucosamine-1-phosphate transferase, alpha and beta subunits | 12q23.2 | 212959_s_at | |
| 11 | PA | KCNMA1 | Potassium large conductance calcium-activated channel, subfamily M, alpha member 1 | 10q22.3 | 221584_s_at | DDX27 | DEAD (Asp-Glu-Ala-Asp) box polypeptide 27 | 20q13.13 | 215693_x_at | 3 |
| | | GNPTAB | N-acetylglucosamine-1-phosphate transferase, alpha and beta subunits | 12q23.2 | 212959_s_at | NDUFS2 | NADH dehydrogenase (ubiquinone) Fe-S protein 2, 49kDa (NADH-coenzyme Q reductase) | 1q23 | 201966_at | |
| | | APOD | Apolipoprotein D | 3q26.2-qter | 201525_at | PPIA | Peptidylprolyl isomerase A (cyclophilin A) | 7p13 | 211378_x_at | |
| | | CD59 | CD59 molecule, complement regulatory protein | 11p13 | 212463_at | SNRPB2 | Small nuclear ribonucleoprotein polypeptide B | 20p12.1 | 202505_at | |
| | | SEMA3E | Sema domain, immunoglobulin domain (Ig), short basic domain, secreted, (semaphorin) 3E | 7q21.11 | 206941_x_at | ADAMTS3 | ADAM metallopeptidase with thrombospondin type 1 motif, 3 | 4q13.3 | 214913_at | |
| | | CD59 | CD59 molecule, complement regulatory protein | 11p13 | 200985_x_at | HINT1 | Histidine triad nucleotide binding protein 1 | 5q31.2 | 208826_x_at | |
| | | BAMBI | BMP and activin membrane-bound inhibitor homolog (Xenopus laevis) | 10p12.13-p11.2 | 203304_at | CIAPIN1 | Cytokine induced apoptosis inhibitor 1 | 16q13-q21 | 208968_s_at | |
| 12 | GBM | FLNA | Filamin A, alpha | Xq28 | 214752_x_at | TNKS2 | Tankyrase, TRF1-interacting ankyrin-related ADP-ribose polymerase 2 | 10q23.3 | 218228_s_at | 1 |
| | | ITGB3BP | Integrin beta 3 binding protein (beta3-endonexin) | 1p31.3 | 205176_s_at | RB1CC1 | RB1-inducible coiled-coil 1 | 8q11 | 202034_x_at | |
| | | DDX27 | DEAD (Asp-Glu-Ala-Asp) box polypeptide 27 | 20q13.13 | 215693_x_at | TRIM8 | Tripartite motif-containing 8 | 10q24.3 | 221012_s_at | |
| 13 | OLG | LARP5 | La ribonucleoprotein domain family, member 4B | 10p15.3 | 208953_at | ANXA1 | Annexin A1 | 9q12-q21.2 | 201012_at | 1 |

**Node #:** Corresponds to numerical labels shown in the brain phenotype diagnostic hierarchy (**Fig. 1**). **Brain phenotype abbreviation (name):** EPN (Ependymoma), GBM (Glioblastoma multiforme), MDL (Medulloblastoma), MNG (Meningioma), normal (Normal brain), OLG (Oligodendroglioma), and PA (Pilocytic astrocytoma). **Gene i / Gene j:** the gene expressed higher and lower in the gene-pair, respectively, within each corresponding phenotype. **Gene name / Chromosome locus:** according to Entrez Gene. **Affymetrix Probe ID:** For both Affymetrix Human Genome U133A and U133Plus2.0 Arrays. **k:** The minimum number of gene-pair classifiers whose decision rule outcomes for a test sample are required to be 'true (= 1)' for the sample to be classified as the phenotype(s) of the corresponding node.

**Table S4.** Decision-tree marker panel for brain cancer and normal transcriptome classification.

| Gene *i* | | Gene *j* | | Brain phenotype binary signature | | | | | | |
|---|---|---|---|---|---|---|---|---|---|---|
| Gene symbol | Affymetrix Probe ID | Gene symbol | Affymetrix Probe ID | EPN | GBM | MDL | MNG | OLG | PA | normal |
| *PRPF40A* | 218053_at | *PURA* | 204021_s_at | 1 | 1 | 1 | 1 | 1 | 1 | 0 |
| *NRCAM* | 204105_s_at | *ISLR* | 207191_s_at | 1 | 1 | 1 | 0 | 1 | 1 | - |
| *SRI* | 208920_at | *NBEA* | 221207_s_at | 1 | 1 | 0 | - | 1 | 1 | - |
| *NUP62CL* | 220520_s_at | *OR10H3* | 208520_at | 1 | 0 | - | - | 0 | 0 | - |
| *DDX27* | 215693_x_at | *KCNMA1* | 221584_s_at | - | 1 | - | - | 1 | 0 | - |
| *FLNA* | 214752_x_at | *TNKS2* | 218228_s_at | - | 1 | - | - | 0 | - | - |

For each classifier decision rule (i.e. Is Gene *i* > Gene *j* ?), 1 and 0 delineates 'true' and 'false', respectively, and ' – ' denotes that the outcome is not used for classification. The vertical binary pattern under each class label corresponds to a phenotype-specific molecular signature.

**Table S5.** Summary of expression differences between genes-pair classifiers.

| Node # | Node phenotype classes | Sample number | Gene *i* Gene symbol | Gene *i* Affymetrix Probe ID | Gene *j* Gene symbol | Gene *j* Affymetrix Probe ID | *k* | Mean | St. dev. | Max. | Min. | Median |
|---|---|---|---|---|---|---|---|---|---|---|---|---|
| 2 | EPN GBM MDL MNG OLG PA | 718 | PRPF40A | 218053_at | PURA | 204021_s_at | 1 | 4,314.7 | 1,685.5 | 10,889.0 | -2,593.0 | 4,370.5 |
| 3 | normal | 203 | PURA | 204021_s_at | PRPF40A | 218053_at | 1 | 1,793.5 | 1,122.2 | 9,249.0 | 176.0 | 1,591.0 |
| 4 | EPN GBM MDL OLG PA | 557 | NRCAM | 204105_s_at | ISLR | 207191_s_at | 1 | 10,376.0 | 2,475.5 | 13,271.0 | -7,491.0 | 11,283.0 |
|  |  |  | IDH2 | 210046_s_at | GMDS | 214106_s_at |  | 5,153.1 | 1,647.1 | 9,264.0 | -6,662.0 | 5,238.0 |
| 5 | MNG | 161 | ISLR | 207191_s_at | NRCAM | 204105_s_at | 1 | 8,568.9 | 2,449.9 | 11,884.0 | -11,674.0 | 8,977.0 |
| 6 | EPN GBM OLG PA | 456 | SALL1 | 206893_at | PAFAH1B3 | 203228_at | 2 | 8,239.6 | 2,767.1 | 12,574.0 | -9,364.0 | 8,728.0 |
|  |  |  | SRI | 208920_at | NBEA | 221207_s_at |  | 4,513.8 | 2,001.8 | 10,692.0 | -401.0 | 4,536.5 |
|  |  |  | DDR1 | 210749_x_at | TIA1 | 201447_at |  | 4,227.7 | 1,528.7 | 9,580.0 | -3,227.0 | 4,536.0 |
|  |  |  | DDR1 | 208779_x_at | MAB21L1 | 206163_at |  | 9,514.9 | 2,223.0 | 13,251.0 | -1,759.0 | 10,267.0 |
|  |  |  | ITPKB | 203723_at | PDS5B | 204742_s_at |  | 7,051.7 | 2,551.4 | 11,373.0 | -1,524.0 | 7,478.5 |
| 7 | MDL | 101 | PAFAH1B3 | 203228_at | SALL1 | 206893_at | 4 | 8,729.6 | 2,776.6 | 12,096.0 | 9.0 | 9,341.0 |
|  |  |  | NBEA | 221207_s_at | SRI | 208920_at |  | 3,858.3 | 2,074.7 | 8,440.0 | -28.0 | 3,715.0 |
|  |  |  | TIA1 | 201447_at | DDR1 | 210749_x_at |  | 3,476.4 | 1,618.5 | 8,709.0 | 27.0 | 3,481.0 |
|  |  |  | MAB21L1 | 206163_at | DDR1 | 208779_x_at |  | 4,510.6 | 1,845.9 | 8,771.0 | -5,415.0 | 4,562.0 |
|  |  |  | PDS5B | 204742_s_at | ITPKB | 203723_at |  | 5,495.4 | 2,515.1 | 11,625.0 | -2,026.0 | 5,538.0 |
| 8 | EPN | 102 | NUP62CL | 220520_at | ZNF280A | 216034_at | 2 | 5,676.0 | 3,444.6 | 11,512.0 | -830.0 | 6,079.5 |
|  |  |  | GALNS | 206335_at | WAS | 38964_r_at |  | 3,952.7 | 2,095.5 | 7,831.0 | -1,216.0 | 4,098.5 |
|  |  |  | CELSR1 | 41660_at | OR10H3 | 208520_at |  | 4,959.8 | 2,798.6 | 11,365.0 | -1,367.0 | 5,058.5 |
|  |  |  | TLE4 | 216997_x_at | OLIG2 | 213824_at |  | 3,423.4 | 2,401.5 | 9,765.0 | -3,239.0 | 3,386.5 |

**Node #:** Corresponds to numerical labels shown in the brain phenotype diagnostic hierarchy (**Fig. 1**). **Brain phenotype abbreviation (name):** EPN (Ependymoma), GBM (Glioblastoma multiforme), MDL (Medulloblastoma), MNG (Meningioma), normal (Normal brain), OLG (Oligodendroglioma), and PA (Pilocytic astrocytoma). **Sample number:** Number of total samples in classes of respective Node #. **Gene *i* / Gene *j*:** the gene expressed higher and lower in the gene-pair, respectively, within each corresponding phenotype. **Gene name / Chromosome locus:** according to Entrez Gene. **Affymetrix Probe ID:** For both Affymetrix Human Genome U133A and U133Plus2.0 Arrays. *k*: The minimum number of gene-pair classifiers whose decision rule outcomes for a test sample are required to be 'true' (= 1) for the sample to be classified as the phenotype(s) of the corresponding node. Ranked expression differences of each gene pair (i.e. Rank_gene_*i* – Rank_gene_*j*) were calculated for each sample, and **Mean, St. dev., Max., Min.,** and **Median** were found across all samples within classes of respective Node #.

**Table S5.** (Continued) Summary of expression differences between genes-pair classifiers.

| Node # | Node phenotype classes | Sample number | Gene *i* | | Gene *j* | | *k* | Mean | St. dev. | Max. | Min. | Median |
|---|---|---|---|---|---|---|---|---|---|---|---|---|
| | | | Gene symbol | Affymetrix Probe ID | Gene symbol | Affymetrix Probe ID | | | | | | |
| 9 | GBM OLG PA | 354 | *ZNF280A* | 216034_at | *NUP62CL* | 220520_s_at | 1 | 1,392.5 | 673.9 | 3,211.0 | -4,279.0 | 1,354.0 |
| 10 | GBM OLG | 292 | *DDX27* | 215693_x_at | *KCNMA1* | 221584_s_at | 1 | 2,144.8 | 1,366.3 | 6,451.0 | -1,681.0 | 2,124.5 |
| | | | *COX7A2* | 217249_x_at | *GNPTAB* | 212959_s_at | | 1,777.2 | 1,106.8 | 5,125.0 | -2,040.0 | 1,664.0 |
| 11 | PA | 62 | *KCNMA1* | 221584_s_at | *DDX27* | 215693_x_at | 3 | 1,832.2 | 942.5 | 3,558.0 | 105.0 | 1,853.0 |
| | | | *GNPTAB* | 212959_s_at | *NDUFS2* | 201966_at | | 1,149.7 | 737.3 | 3,061.0 | -825.0 | 1,112.5 |
| | | | *APOD* | 201525_at | *PPIA* | 211378_x_at | | 121.4 | 117.3 | 494.0 | -426.0 | 104.5 |
| | | | *CD59* | 212463_at | *SNRPB2* | 202505_at | | 1,169.1 | 1,038.0 | 3,989.0 | -2,247.0 | 1,059.0 |
| | | | *SEMA3E* | 206941_x_at | *ADAMTS3* | 214913_at | | 4,376.3 | 3,026.1 | 8,551.0 | -8,061.0 | 5,194.5 |
| | | | *CD59* | 200985_s_at | *HINT1* | 208826_x_at | | 388.5 | 422.5 | 1,586.0 | -270.0 | 214.5 |
| | | | *BAMBI* | 203304_at | *CIAPIN1* | 208968_s_at | | 1,976.4 | 975.1 | 3,802.0 | -594.0 | 1,984.0 |
| 12 | GBM | 231 | *FLNA* | 214752_x_at | *TNKS2* | 218228_s_at | 1 | 3,238.8 | 2,804.3 | 9,691.0 | -4,133.0 | 3,557.0 |
| | | | *ITGB3BP* | 205176_s_at | *RB1CC1* | 202034_x_at | | 1,793.7 | 2,357.9 | 8,665.0 | -4,597.0 | 1,560.0 |
| | | | *DDX27* | 215693_x_at | *TRIM8* | 221012_s_at | | 1,819.1 | 1,492.6 | 6,083.0 | -1,099.0 | 1,636.0 |
| 13 | OLG | 61 | *LARP5* | 208953_at | *ANXA1* | 201012_at | 1 | 2,843.4 | 3,463.0 | 9,055.0 | -5,831.0 | 2,788.0 |

**Node #:** Corresponds to numerical labels shown in the brain phenotype diagnostic hierarchy (**Fig. 1**). **Brain phenotype abbreviation (name):** EPN (Ependymoma), GBM (Glioblastoma multiforme), MDL (Medulloblastoma), MNG (Meningioma), normal (Normal brain), OLG (Oligodendroglioma), and PA (Pilocytic astrocytoma). **Sample number:** Number of total samples in classes of respective Node #. **Gene *i* / Gene *j*:** the gene expressed higher and lower in the gene-pair, respectively, within each corresponding phenotype. **Gene name / Chromosome locus:** according to Entrez Gene. **Affymetrix Probe ID:** For both Affymetrix Human Genome U133A and U133Plus2.0 Arrays. ***k*:** The minimum number of gene-pair classifiers whose decision rule outcomes for a test sample are required to be 'true (= 1)' for the sample to be classified as the phenotype(s) of the corresponding node. Ranked expression differences of each gene pair (i.e. Rank_gene_*i* − Rank_gene_*j*) were calculated for each sample, and **Mean, St. dev., Max., Min.,** and **Median** were found across all samples within classes of respective Node #.

**Table S6.** Hold-one-lab-in validation accuracies of glioblastoma signatures.

| GBM training set (sample size) | GBM test set (sample size) | Predicted phenotype / % of test set / samples of test set |
| --- | --- | --- |

**GSE4412 (59)**

GSE4271 (76)

| | UC | EPN | GBM | MDL | MNG | OLG | PA | Total |
| --- | --- | --- | --- | --- | --- | --- | --- | --- |
| | 2.63% | 57.89% | **9.21%** | 17.11% | 5.26% | 1.32% | 6.58% | 76 |
| | 2 | 44 | 7 | 13 | 4 | 1 | 5 | 76 |

GSE8692 (6)

| | GBM | MNG | Total |
| --- | --- | --- | --- |
| | **83.33%** | 16.67% | 6 |
| | 5 | 1 | 6 |

GSE9171 (13)

| | EPN | GBM | MNG | Total |
| --- | --- | --- | --- | --- |
| | 92.31% | **0.00%** | 7.69% | 13 |
| | 12 | 0 | 1 | 13 |

GSE4290 (77)

| | EPN | GBM | MDL | MNG | PA | Total |
| --- | --- | --- | --- | --- | --- | --- |
| | 85.71% | **0.00%** | 2.60% | 5.19% | 6.49% | 77 |
| | 66 | 0 | 2 | 4 | 5 | 77 |

**GSE4271 (76)**

GSE4412 (59)

| | UC | GBM | PA | normal | Total |
| --- | --- | --- | --- | --- | --- |
| | 11.86% | **77.97%** | 8.47% | 1.69% | 59 |
| | 7 | 46 | 5 | 1 | 59 |

GSE8692 (6)

| | GBM | Total |
| --- | --- | --- |
| | **100.0%** | 6 |
| | 6 | 6 |

GSE9171 (13)

| | GBM | 6 | Total |
| --- | --- | --- | --- |
| | **92.31%** | 7.69% | 13 |
| | 12 | 1 | 13 |

GSE4290 (77)

| | UC | GBM | MNG | PA | Total |
| --- | --- | --- | --- | --- | --- |
| | 5.19% | **77.92%** | 1.30% | 15.58% | 77 |
| | 4 | 60 | 1 | 12 | 77 |

**GSE8692 (6)**

GSE4412 (59)

| | UC | EPN | GBM | MDL | MNG | PA | normal | Total |
| --- | --- | --- | --- | --- | --- | --- | --- | --- |
| | 5.08% | 13.56% | **47.46%** | 1.69% | 3.39% | 27.12% | 1.69% | 59 |
| | 3 | 8 | 28 | 1 | 2 | 16 | 1 | 59 |

GSE4271 (76)

| | UC | EPN | GBM | MDL | PA | normal | Total |
| --- | --- | --- | --- | --- | --- | --- | --- |
| | 9.21% | 32.89% | **18.42%** | 5.26% | 32.89% | 1.32% | 76 |
| | 7 | 25 | 14 | 4 | 25 | 1 | 76 |

GSE9171 (13)

| | EPN | GBM | MDL | PA | Total |
| --- | --- | --- | --- | --- | --- |
| | 61.54% | **15.38%** | 15.38% | 7.69% | 13 |
| | 8 | 2 | 2 | 1 | 13 |

GSE4290 (77)

| | UC | EPN | GBM | MDL | MNG | PA | normal | Total |
| --- | --- | --- | --- | --- | --- | --- | --- | --- |
| | 14.29% | 42.86% | **7.79%** | 1.30% | 1.30% | 25.97% | 6.49% | 77 |
| | 11 | 33 | 6 | 1 | 1 | 20 | 5 | 77 |

**Table S6.** (Continued) Hold-one-lab-in validation accuracies of glioblastoma signatures.

**GSE9171 (13)**

GSE4412 (59)

| UC | EPN | GBM | MDL | MNG | PA | Total |
|---|---|---|---|---|---|---|
| 35.59% | 13.56% | **0.00%** | 1.69% | 5.08% | 44.07% | 59 |
| 21 | 8 | 0 | 1 | 3 | 26 | 59 |

GSE4271 (76)

| UC | EPN | GBM | MDL | MNG | PA | Total |
|---|---|---|---|---|---|---|
| 19.74% | 38.16% | **0.00%** | 6.58% | 3.95% | 31.58% | 76 |
| 15 | 29 | 0 | 5 | 3 | 24 | 76 |

GSE8692 (6)

| UC | GBM | MNG | PA | Total |
|---|---|---|---|---|
| 66.67% | **0.00%** | 16.67% | 16.67% | 6 |
| 4 | 0 | 1 | 1 | 6 |

GSE4290 (77)

| UC | EPN | GBM | MDL | PA | normal | Total |
|---|---|---|---|---|---|---|
| 10.39% | 40.26% | **0.00%** | 1.30% | 46.75% | 1.30% | 77 |
| 8 | 31 | 0 | 1 | 36 | 1 | 77 |

**GSE4290 (77)**

GSE4412 (59)

| UC | GBM | NB | PA | normal | Total |
|---|---|---|---|---|---|
| 5.08% | **52.54%** | 27.12% | 13.56% | 1.69% | 59 |
| 3 | 31 | 16 | 8 | 1 | 59 |

GSE4271 (76)

| UC | EPN | GBM | MDL | OLG | PA | Total |
|---|---|---|---|---|---|---|
| 1.32% | 1.32% | **60.53%** | 3.95% | 15.79% | 17.11% | 76 |
| 1 | 1 | 46 | 3 | 12 | 13 | 76 |

GSE8692 (6)

| UC | GBM | NB | PA | Total |
|---|---|---|---|---|
| 33.33% | **16.67%** | 16.67% | 33.33% | 6 |
| 2 | 1 | 1 | 2 | 6 |

GSE9171 (13)

| UC | GBM | Total |
|---|---|---|
| 7.69% | **92.31%** | 13 |
| 1 | 12 | 13 |

**Table S7.** Hold-one-lab-in (H1LI) and leave-one-lab-out (L1LO) validation accuracies of glioblastoma signatures when training data were constrained to 50 total samples.

| Method | GBM training set (50 samples) | GBM test set | GBM prediction | | Average performance |
|---|---|---|---|---|---|
| | | | Average accuracy | St. dev. | |
| H1LI | GSE4412 | GSE4271 | 40.26% | 14.98% | 36.39% |
| | | GSE8692 | 96.67% | 7.03% | |
| | | GSE9171 | 6.15% | 3.24% | |
| | | GSE4290 | 2.47% | 2.10% | |
| | GSE4271 | GSE4412 | 58.98% | 21.64% | 63.89% |
| | | GSE8692 | 74.00% | 11.43% | |
| | | GSE9171 | 73.08% | 10.41% | |
| | | GSE4290 | 49.48% | 26.56% | |
| | GSE4290 | GSE4412 | 38.47% | 10.23% | 40.66% |
| | | GSE4271 | 43.13% | 16.12% | |
| | | GSE8692 | 23.33% | 9.08% | |
| | | GSE9171 | 57.70% | 16.79% | |
| L1LO | GSE4271, GSE8692, GSE9171, GSE4290 | GSE4412 | 82.20% | 10.39% | 69.72% |
| | GSE4412, GSE8692, GSE9171, GSE4290 | GSE4271 | 54.87% | 7.18% | |
| | GSE4412, GSE4271, GSE8692, GSE9171 | GSE4290 | 72.08% | 15.29% | |

H1LI and L1LO validations were performed ten times for each category of training data. In each validation trial, 50 samples were randomly selected from the single microarray dataset (for H1L1) or from the multi-study, combined dataset (for L1LO).

**Table S8.** Ten-fold cross-validation accuracies when only the node marker panel was required to reach unique diagnoses.

| Phenotype | Total samples | Sample size (%) | Accuracy (%) |
|-----------|---------------|-----------------|--------------|
| EPN | 102 | 93.1 | 95.8 |
| GBM | 231 | 88.9 | 92.7 |
| MDL | 101 | 95.0 | 95.8 |
| MNG | 161 | 98.8 | 97.5 |
| OLG | 61 | 77.0 | 74.5 |
| PA | 62 | 90.3 | 96.4 |
| Normal | 203 | 97.9 | 99.5 |
| Average | - | 91.6 | 93.2 |

**Sample size:** Average proportion of total samples that reached unique diagnoses via node marker panel. **Accuracy:** Reflects average performance in ten-fold cross-validation conducted ten times.

**Table S9.** Functional roles of 11 previously identified GBM serum markers that are present in our extracellular-product encoding marker-panel.

| Gene | Gene name | Functional role[a] |
|------|-----------|--------------------|
| *APOD* | Apolipoprotein D | This gene encodes a component of high density lipoprotein that has no marked similarity to other apolipoprotein sequences. It has a high degree of homology to plasma retinol-binding protein and other members of the alpha 2 microglobulin protein superfamily of carrier proteins, also known as lipocalins. This glycoprotein is closely associated with the enzyme lecithin:cholesterol acyltransferase - an enzyme involved in lipoprotein metabolism. |
| *CALU* | Calumenin | The product of this gene is a calcium-binding protein localized in the endoplasmic reticulum (ER) and it is involved in such ER functions as protein folding and sorting. This protein belongs to a family of multiple EF-hand proteins (CERC) that include reticulocalbin, ERC-55, and Cab45 and the product of this gene. Alternatively spliced transcript variants encoding different isoforms have been identified. |
| *CD163* | Cluster of Differentiation 163 | The protein encoded by this gene is a member of the scavenger receptor cysteine-rich (SRCR) superfamily, and is exclusively expressed in monocytes and macrophages. It functions as an acute phase-regulated receptor involved in the clearance and endocytosis of hemoglobin/haptoglobin complexes by macrophages, and may thereby protect tissues from free hemoglobin-mediated oxidative damage. This protein may also function as an innate immune sensor for bacteria and inducer of local inflammation. Alternatively spliced transcript variants encoding different isoforms have been described for this gene. |
| *CHI3L1* | Chitinase-3-like protein 1 | Chitinases catalyze the hydrolysis of chitin, which is an abundant glycopolymer found in insect exoskeletons and fungal cell walls. The glycoside hydrolase 18 family of chitinases includes eight human family members. This gene encodes a glycoprotein member of the glycosyl hydrolase 18 family. The protein lacks chitinase activity and is secreted by activated macrophages, chondrocytes, neutrophils and synovial cells. The protein is thought to play a role in the process of inflammation and tissue remodeling. |
| *CSF1* | Colony stimulating factor 1 | The protein encoded by this gene is a cytokine that controls the production, differentiation, and function of macrophages. The active form of the protein is found extracellularly as a disulfide-linked homodimer, and is thought to be produced by proteolytic cleavage of membrane-bound precursors. The encoded protein may be involved in development of the placenta. Alternate splicing results in multiple transcript variants. |
| *EGFR* | Epidermal growth factor receptor | The protein encoded by this gene is a transmembrane glycoprotein that is a member of the protein kinase superfamily. This protein is a receptor for members of the epidermal growth factor family. EGFR is a cell surface protein that binds to epidermal growth factor. Binding of the protein to a ligand induces receptor dimerization and tyrosine autophosphorylation and leads to cell proliferation. Mutations in this gene are associated with lung cancer. Multiple alternatively spliced transcript variants that encode different protein isoforms have been found for this gene. |
| *IGFBP2* | Insulin-like growth factor binding protein 2 | Inhibits IGF-mediated growth and developmental rates. IGF-binding proteins prolong the half-life of the IGFs and have been shown to either inhibit or stimulate the growth promoting effects of the IGFs on cell culture. They alter the interaction of IGFs with their cell surface receptors. |
| *NID1* | Nidogen 1 | This gene encodes a member of the nidogen family of basement membrane glycoproteins. The protein interacts with several other components of basement membranes, and may play a role in cell interactions with the extracellular matrix. |
| *PDGFC* | Platelet derived growth factor C | The protein encoded by this gene is a member of the platelet-derived growth factor family. The four members of this family are mitogenic factors for cells of mesenchymal origin and are characterized by a core motif of eight cysteines. This gene product appears to form only homodimers. It differs from the platelet-derived growth factor alpha and beta polypeptides in having an unusual N-terminal domain, the CUB domain. Alternatively spliced transcript variants have been found for this gene. |
| *PSG9* | Pregnancy specific beta-1-glycoprotein 9 | The human pregnancy-specific glycoproteins (PSGs) are a group of molecules that are mainly produced by the placental syncytiotrophoblasts during pregnancy. PSGs comprise a subgroup of the carcinoembryonic antigen (CEA) family, which belongs to the immunoglobulin superfamily. |
| *PTN* | Pleiotrophin | Secreted growth factor that induces neurite outgrowth and which is mitogenic for fibroblasts, epithelial, and endothelial cells. Binds anaplastic lymphoma kinase (ALK) which induces MAPK pathway activation, an important step in the anti-apoptotic signaling of PTN and regulation of cell proliferation |

[a]http://www.ncbi.nlm.nih.gov/gene

**Table S10.** Ten-fold cross-validation accuracies of gene-pair classifiers composed of genes that encode extracellular products.

| | | Predicted phenotype (%)[a] | | | | | | | | total |
|---|---|---|---|---|---|---|---|---|---|---|
| | | EPN | GBM | MDL | MNG | OLG | PA | normal | UC[b] | |
| Actual phenotype | EPN | **83.3** | 2.9 | 4.9 | 4.9 | 2.0 | 1.0 | 0.0 | 1.0 | 102 |
| | GBM | 2.2 | **72.3** | 0.4 | 1.3 | 20.3 | 0.9 | 0.0 | 2.6 | 231 |
| | MDL | 1.0 | 1.0 | **94.1** | 1.0 | 2.0 | 0.0 | 0.0 | 1.0 | 101 |
| | MNG | 0.0 | 1.2 | 0.6 | **97.5** | 0.0 | 0.0 | 0.0 | 0.6 | 161 |
| | OLG | 0.0 | 16.4 | 0.0 | 0.0 | **75.4** | 3.3 | 1.6 | 3.3 | 61 |
| | PA | 1.6 | 1.6 | 0.0 | 0.0 | 3.2 | **91.9** | 0.0 | 1.6 | 62 |
| | normal | 0.0 | 1.0 | 0.0 | 0.0 | 1.5 | 0.0 | **96.6** | 1.0 | 203 |

[a]Accuracies reflect average performance in ten-fold cross-validation conducted ten times. The main diagonal gives the average classification accuracy of each class (bold), and the off-diagonal elements show the erroneous predictions.
[b]UC (Unclassified samples). When using the node classifiers, expression profiles that did not exert a signature of any phenotype (i.e., did not percolate down to at least one positive terminal node) were rejected from classification. In this case, the Unclassified sample is treated as a misclassification.